\documentclass{emulateapj}
\usepackage{natbib,graphicx,psfig}

\def\refnew#1{(\ref{#1})}
\def \micron {\, \mu {\rm m}}
\def \eV {\, \rm eV}
\def \AU {\, \rm AU}
\def \cm {\, \rm cm}
\def \mm {\, \rm mm}
\def \g {\, \rm g}
\def \s {\, \rm s}
\def \K {\, \rm K}
\def \erg {\, \rm erg}

\def \be {\begin{equation}}
\def \ee {\end{equation}}

\begin{document}

%\title{Ring Formation in Debris Disks around A-type Stars}
\title{Formation of Narrow Dust Rings in Circumstellar Debris Disks}
%Axi-Symmetric Ring Structure in Circumstellar Disks about A-Type Stars}

\author{Gurtina Besla\altaffilmark{1,2} and Yanqin Wu\altaffilmark{2}}
\altaffiltext{1}{Harvard-Smithsonian Center for Astrophysics, 60 Garden Street, 
Cambridge, MA 02138, USA}
\altaffiltext{2}{Department of Astronomy \& Astrophysics, 50 St. George Street, University of Toronto,
    Toronto, ON M5S 3H4, Canada}
\email{gbesla@cfa.harvard.edu,wu@astro.utoronto.ca}

\begin{abstract}
Narrow dust rings observed around some young stars (e.g., HR 4796A)
% and HD 105) 
need to be confined. We present a possible explanation for the
formation and confinement of such rings in optically thin
circumstellar disks, without invoking shepherding planets.
% condition for progressively thinner...
If an enhancement of dust grains (e.g., due to a catastrophic
collision) occurs somewhere in the disk, photoelectric emission from
the grains can heat the gas to temperatures well above that of the
dust. The gas orbits with super(sub)-Keplerian speeds inward (outward) 
of the associated pressure maximum. This tends to concentrate the
grains into a narrow region. The rise in dust density leads to further
heating and a stronger concentration of grains.
%more concentration. 
A narrow dust ring forms as a result of this instability.
%Consequently, the dust ring narrows progressively.
% A progressively narrower dust ring forms.
%despite grain-grain collision that tend to spread...
%
We show that this mechanism not only operates around early-type stars
that have high UV fluxes, but also around stars with spectral types as
late as K. This implies that this process is generic and may have
occurred during the lifetime of each circumstellar disk.
% 
%
%generic in terms of masses... grain-grain collision
%
%This mechanism requires a dust-to-gas mass-ratio of order unity and is
%most effective around A-type stars due to their higher UV fluxes. It
%may explain the formation of narrow rings as observed in debris disks
%around HR 4796A and other objects.
%
%
%Our analysis builds on the theory first proposed by
%\citet{TA} 
%but does not require an artificial sharp edge to the
%%gas-disk, 
%and it complements the linear stability analysis recently published by
%\citet{Klahr}. However, subsequent evolution of the dust ring as the gas 
%dissipates deserves further study.
%
We examine the stringent upper-limit on the H$_2$ column density in 
the HR 4796A disk and find it to be compatible with the presence of a
significant amount of hydrogen gas in the disk.
We also compute the OI and CII infrared line fluxes expected from
various debris disks and show that these will be easily detectable by
the upcoming Herschel mission. Herschel will be instrumental in
detecting and characterizing gas in these disks.
%
%
%Instead, we assume that a broad (FWHM of 20 AU), Kuiper-belt-like dust
%enhancement exists at 70 AU (total mass $M_{\rm {dust}} = M_{\rm
%{gas}} = 0.1M_\oplus$). We focus our analysis on the sharp ring
%structure (width $\lesssim$ 17 AU) observed about the A0V star,
%HR4796A \citep{Schneider}. We rigorously demonstrate that the
%described local dust enhancement will increase the magnitude of
%dust-dependent heating rates, such as photoelectric heating, and
%consequently dramatically increase the local gas temperature. The
%corresponding change in gas pressure counteracts the tendency for
%grains $\gtrsim$ 16$\mu$m to migrate outwards due to radiation
%pressure. Instead, these grains migrate to stable orbits defined by
%grain size, producing a single, narrow (FWHM $\lesssim$ 10 AU) ring
%structure centred at 70 AU. The resulting increase in local dust
%density further heats the gas such that smaller, collisionally
%regenerated grains can be trapped - i.e. the narrowing process is
%self-perpetuating. The effectiveness of this mechanism is dependent on
%the dominance of the photoelectric heating rate and the relative
%weakness of heating/cooling via gas-grain collisions. Consequently,
%the predicted dusty ring is more pronounced in gas-depleted disks
%about central stars with a strong UV flux, such as HR4796A; however,
%the subsequent evolution of the ring feature as the gas slowly
%dissipates deserves further study.
\end{abstract}


\section{Introduction}
\label{sec:intro}

Observations using infrared space telescopes, such as the {\it
Infrared Space Observatory} ({\it ISO}), the {\it Infrared
Astronomical Satellite} ({\it IRAS}) and recently the {\it Spitzer
Space Telescope}, have demonstrated that many stars exhibit the
{\it``Vega-phenomenon,''} --- excess infrared radiation by orbiting
dust grains heated by the central star
\citep{Zuckerman}. These grains are believed to lie in
optically-thin, gas-poor disks that are likely the remains of more
massive protoplanetary disks
\citep{Backman,SE}. Such disks are termed `debris disks' 
because the grains have short life-times and must be continuously
regenerated by larger bodies.  Famous examples of this phenomenon
include Vega \citep{Aumann84} and $\beta$ Pic \citep[][and references
therein]{Aumann85}. \citet{MeyerPPV} presents the most recent review
of debris disk observations.

When spatially resolved, many of these disks display peculiar
morphologies, including rings, holes, clumps and warps. Properties of
some well-known debris disks and their host stars are summarized in
Table~\ref{tbl-1}. Ring-like features appear to be prevalent. In this
article, we focus on the phenomenology of narrow rings, with the two
most prominent examples being disks around the two A-type stars HR
%GB AUG 31 changed the order - Schneider last  and Marsh last
4796A \citep{RayJay,Koerner98,Schneider} and Fomalhaut
\citep{Holland,kalas2,Marsh}. In the former case, 
a dust ring is seen in scattered light by HST at a distance of 70 AU
away from the star with an un-resolved width $< 17$ AU
\citep{Schneider}. The dust disk in Fomalhaut was resolved by 
HST to be distributed in a narrow belt (width $\sim 25$ AU) at a 
distance of 135 AU from the star \citep{kalas2}. 

% added two new systems
Narrow dust rings also appear around stars of later spectral types:
the dust disks of two F-type stars, HD 139664 (F5) and HD 181327
(F5/F6) were recently resolved by HST as narrow belts, the former
%GB AUG 31 : centering to centered 
% 83 AU changed to 86 AU - Schneider 2006 paper
appears as a ring of width $\sim 26$ AU centered $\sim 90$ AU away
from the star \citep{Kalas2006}, while the latter a width $36$ AU at
$\sim 86$ AU away \citep{Schneider2006}.  In addition, model fitting
using Spitzer spectrophotometry data indicate that dust around the
solar-type star HD 105 may lie in a narrow belt (width $< 4$ AU)
located $\sim42$ AU away from the star \citep{Meyer}.
% Gurtina Aug 3:  reworded the below:
%Narrow dust rings also appear around solar-type stars:
%Resolved images
%by HST revealed a broad ring (width $\sim 85$ AU) around the G2 star
%HD 107146 \citep{Ardila}, while 
%model fitting using Spitzer spectrophotometry data indicate that dust
%around the G0 star HD 105 may lie in a narrow belt (width $< 4$ AU)
%located $\sim42$ AU away from the star \citep{Meyer}. 

While HR 4796A, HD 181327
% this above is inconclusive, depending on model fitting
and HD 105 are relatively young (a few Myrs), the ages for Fomalhaut and
HD 139664 are estimated to be $\sim 10^8$ years.
%Clumpy, strongly asymmetric rings have been detected in disks around
%older stars like $\epsilon$ Eridani
%\citep[K2][]{Greaves} and Vega [A0][]\citep{Koerner, Wilner}.\footnote{
%Intriguingly, while the Vega disk appears to be clumpy in
%submilli-meter, it is observed in infread to `be smooth with an inner
%hole \citep{Su}.}  
%As such, the possibility that rings/inner holes are
%intrinsic evolutionary phenomena, perhaps common to stars of all
%spectral types, cannot be easily dismissed.
%

%The origin of these narrow rings is puzzling. Even if dust is
%initially generated in a narrow annulus, radiation pressure,
%grain-grain collisions and Poynting-Robertson drag all tend to
%disperse the dust radially \citep[for a discussion,
%see][]{augereau}. In particular, radiation pressure launches small
%grains into orbits of various eccentricities and should disperse the
%narrow ring within a dynamical timescale. Thus, the presence of
%sharp-edged narrow rings implies grain confinement.
%
%wu: new paragraphs (commented out the old one)
Barring issues related to observational sensitivity limits, the
presence of a clear outer edge for these rings is puzzling.  The
surface brightness of debris disks is likely dominated by grains just
above the radiative blow-out size ($\beta=1/2$, see
eq. [\ref{eq6}]). Even if these grains are produced by parent bodies
on circular orbits that lie within a ring, radiation pressure launches
them to eccentric orbits upon creation (with $e =
\beta/(1-\beta)$). The narrow ring geometry is destroyed within a
dynamical time and one observes a greatly broadened disk with no sharp
outer edge \citep[][also see \S
\ref{subsec:eccentric}]{augereau,Thebault,strubbe,Krivov2006,Wyatt2006}.
%This likely explains
%the observed distributions of grain populations in systems such as AU
%Mic, $\beta$ Pic and Vega: small grains are observed in a broad disk,
%whereas larger grains (in the latter two cases) are observed in
%ring-like geometry. 
% Gurtina Aug 3 :  reworded the above to include the Wyatt work.
%wu: the Wyatt 2006 work doesn't seem relevant
This likely explains the distribution of grain populations in AU Mic,
$\beta$ Pic and Vega: small grains are observed in a broad disk,
whereas larger grains (detected in thermal infrared in the cases of
$\beta$ Pic and Vega) are observed in ring-like geometry.
So why do some systems appear ring-like in scattered-light images?
%However, viewed in scattered light, why do some disks appear ring-like
%and some disks appear broad is the question we are after.

% Gurtina Aug 3:  Added 3 new systems. Still missing some info

%wyq: remove the 'omega' column in the table.
% wyq: this table still needs work ?????????????
%	also put in s_min
%	references should be cut to have only relevant ones
% 	only cite resolved systems
% didn't have time to merge this with Gurtina's one yet ?????
\begin{deluxetable*}{ccccccccccc}
\tabletypesize{\scriptsize}
%\rotate
%wu: removed 'and their host stars'
\tablecaption{Properties of some spatially resolved debris disks}
%with Ring-like Features.}
\tablewidth{0pt}
\tablehead{
\colhead{Star} & \colhead{Spectral} 
& \colhead{$L_\ast$} & \colhead{Age} &
\colhead{$\chi$\tablenotemark{a}} &
%\colhead{$\omega_{\rm Si}$\tablenotemark{b}} &
%\colhead{$\omega_{\rm C}$\tablenotemark{c}} & 
\colhead{$M_{\rm dust}$\tablenotemark{b}} &
%wu: changed FWHM --> Ring Width
%wu: changed peak mid-IR --> Peak
\colhead{$L_{\rm IR}$} & \colhead{Peak} & \colhead{Ring Width} &
 \colhead{Refs.} \\
\colhead{} &\colhead{Type} 
& \colhead{($L_\odot$)} & \colhead{(Myr)} & 
%\colhead{} & \colhead{}
\colhead{} & \colhead{($M_\oplus$)} & \colhead{$(L_\ast)$} &
\colhead{emission (AU)} & \colhead{(AU)} & \colhead{} }
\startdata
%wyq: HR < 17AU, optical scattered light
HR 4796A & A0 & 21 & 8 $\pm$2 & 538 & 
%$10^4$& $10^5$&  
$\ge$0.25 & $5\times 10^{-3}$ & $70$ & $< 17$ & 1,2,3,4 \\
%wyq: Vega data from submm & spitzer
Vega & A0 & 60 & $350$ & 
%\nodata & \nodata & 
\nodata & 0.003 &
$2\times 10^{-5}$ & $> 86$ & \nodata & 5,6,7,8 \\
% wyq: changed Vega to a disk > 86AU, only large grains is assumed to lie
%	between 86 and 200AU
% wyq: Fomalhaut width 25AU around 135AU, data from HST & submm
Fomalhaut & A3 & 13 & 100-300 & 
%\nodata & \nodata & 
\nodata & 0.017 &
$4.6\times 10^{-5}$ & $135$ & $25$ & 6,9,12,13 \\
% wyq: changed beta-Pic, scattered light data, 
$\beta$ Pic & A5 & 8.7 & 12$^{+8}_{-4}$ & 0.17 & 
%$10^3$ & $10^7$ &
$0.04$ & $3\times 10^{-3}$ & $120$ & {$90$\tablenotemark{c}} & 10,12,15 \\
%Gurtina Aug 3: added HD 139664, took Lir/L* as HST data
%NO IDEA WHERE THE AGE DETERMINATION CAME FROM
HD 139664 & F5 & 3.3 & $300^{+700}_{-200}$ & \nodata & $>
8.7\times10^{-4}$ & $10^{-5}$ & $\sim 90$ & 26 & 16 \\
% Gurtina Aug 3: added HD 181327
HD 181327 & F5/F6 & 3.1  & $\sim10$ & \nodata & 0.05 & $2\times 10^{-3}$ & 86 & 36 & 17,18 \\
% wyq: added HD 105, distance at 40pc, M_v = 7.53,
HD 105 & G0 & 1.3 & $30\pm 10$ & \nodata  & $\sim 0.03$ &  $3.9\times 10^{-4}$ & 42 & 4 & 19\\
% 
% width 85AU for the following object, HST & submm
HD 107146 & G2 & 1.1 & 80-200 & $3.2\times10^{-6}$& 
%$10^4$ & $10^9$ &
0.1-0.4 & $1.2\times 10^{-3}$ & $130$ & $85$ &5,11 \\
% Gurtina Aug 3: Added HD 53143 - but not sure if should keep
HD 53143 & K1 & 0.7 & $1000\pm200$ & \nodata & $1.2\times 10^{-4}$ & $> 1.6\times10^{-5}$ & $\sim55-110$ & $>55$ & 7,16\\ 
%wyq: changed Eridani according to Greaves05, submm resolved
$\epsilon$ Eridani & K2 & 0.33 & $850$ & 
%\nodata & \nodata & 
\nodata & 0.002 &$8\times 10^{-5}$ & $65$ & $50$ & 12,14 \\ 
\enddata
\tablenotetext{a}{The stellar FUV ($11.2$ - $12.42$ eV) photon
flux, normalized by the average interstellar value of $1.2 \times
10^7/\cm^2/\s$
\citep{habing}, and evaluated at 70 AU from the
star.  The high resolution spectra are provided by Kamp \& Hauschildt
\citep{Hauschildt}, and the flux calibration is performed by
\citet{Fernandez}.}
% gb : changed the eV range to be consistent with Rodrigo's calculations
%wyq: changed the ev range, from 912 to 1110 ang, see Kamp & van Zadelhoff '01
% number flux is 1.2x10^7/cm^2/s, energy flux $2.4 \times 10^{-4}$ erg/cm$^2$/s
\tablenotetext{b}{Dust masses are estimated from sub-millimeter fluxes 
assuming optically thin radiation, except for HD 139664 and HD 53143 
for which the dust masses were calculated from the optical depth as 
described in [16].}
\tablenotetext{c}{We adopt here the radial profile 
inversion by Artymowicz (private communication) based on observations
from \citet{heap}.}
\tablerefs{
(1) \citet{Jura98}; (2) \citet{Stauffer}; (3) \citet{Greaves2}; (4)
\citet{Schneider}; (5) \citet{Williams}; (6) \citet{Dent};
(7) \citet{Song}; (8) \citet{Su}; (9) \citet{Barrado}; (10)
\citet{Zuck2001}; (11) \citet{Ardila}; (12) \citet{DiFolco}; (13) 
\citet{kalas2}; (14) \citet{greaves05}; (15) \citet{heap};
(16) \citet{Kalas2006}; (17) \citet{Schneider2006}; 
(18) \citet{Mannings}; (19) \citet{Meyer}.}
\label{tbl-1}
\end{deluxetable*}

% gurtina Aug 3 : not sure if this is the best place for the Kalas 2006 
% reference, but they specifically address the issue of confinement as 
% the reason for the bifurcation in disk architecture (ring vs broad disk)
%{\bf As noted by \citet{Kalas2006}, without a confinement mechanism for the 
%outer radius, the dust distribution will manifest as a broad disk.} 
%wu: remove. Talked about it above the table (they are just copying others)
The most popular explanation for ring confinement invokes one or more
massive shepherding planets orbiting at distances comparable to that
of the ring.  As grains spiral inward due to Poynting-Robertson (P-R)
drag, the gravitational tug from the planets may temporarily stall the
grains at special radii (mean-motion resonances, MMR), giving rise to
an apparent narrow ring
\citep[see, e.g.][]{Ozernoy, Wilner, Quillen, kuchner,Moran}.
The outward migration of a large planet may also trap grains in MMRs,
without the need for P-R drag \citep[see,
e.g.][]{Wyatt2003,Wyatt2006}. This process is similar to the resonance
capture of Kuiper Belt objects by Neptune. Alternatively,
gravitational scattering by massive planets may prevent grains from
migrating across the planet's orbit, leading to a dust distribution
with a sharp inner edge \citep{moromartin}.
%There also have been work looking into the effects of external stars
%\citep{wyatt99,augereau2}. 

%wu: new paragraph
However, there are problems with the planet hypothesis, even if we
accept the existence of giant planets at such large radii.  Firstly,
grain-grain collisions typicaly induce radial diffusion at a rate
orders of magnitude above that from P-R drag, thus weakening the
capture in MMRs. The sharp edges formed by planet scattering may also
be smeared out by collisions.  Secondly, the ability of planets to
confine grains that are launched to elliptic orbits may be limited.
% Gurtina Aug 3 - not sure if the moran comment is necessary.
\citet{Moran} noted that at eccentricities larger 
than that of the planet's orbit, grains spiralling inwards due to P-R
drag can no longer be trapped in MMRs. 
%For the $\epsilon$ Eridani system, this corresponds
% to an eccentricity of $\sim0.6$, although the resonant structures begin to 
%dissipate at eccentricities as low as $\sim0.4$. 
% this is key, Wyatt 2006 results:
\citet{Wyatt2006} further illustrated that even grains generated by a
resonant parent population will become radially spread out and
azimuthally axisymmetric, so long as they are small enough to
experience significant radiation pressure ($\beta \sim 1/2$).
%Furthermore, in the planet migration model, \citet{Wyatt2006}
%illustrated that collisionally regenerated grains launched into
%elliptic orbits will segregate into 3 grain populations with distinct
%spatial structures: the smallest grains (with $\beta < 0.5$) are blown
%out of the system and follow a density distribution that falls off as
%$\tau \alpha r^{-1}$; large grains (with $\beta > \beta_{\rm crit}$,
%where $\beta_{\rm crit}$ is a function of the planet mass) remain
%within the same resonant structures as the planetesimals; the third
%class of particles (with $\beta_{\rm crit} < \beta < 0.5$) are no
%longer in resonance and exhibit a radially extended
%distribution. These distinct grain populations are consistent with
%observations of Vega, but not of systems such as HR
%4796A. Specifically, a sharp outer edge cannot be maintained in this
%scenario. 
Further studies taking these effects into account are clearly
warranted.
% Gurtina Aug 3: can't say the below - eccentricity has been examined.
%Firstly,
%the ability of planets to confine grains that are launched to elliptic
%orbits has not been investigated.
% Secondly, grain-grain collisions typically induce radial diffusion at a 
%rate orders of magnitude above that from P-R drag, thus weakening the 
%trapping by MMR. 
%The sharp edges formed by planet scattering may also be smeared out by 
%collisions. Further studies taking these two
%effects into account are clearly warranted.
%Firstly, it remains difficult to envision the ubiquitous presence of massive
%planets at such large distances. Secondly, the effects of grain-grain
%collisions have been ignored in the above mentioned works. 
%For the observed values of grain number density, 
%If instead planets, acting as shepherding bodies, dynamically
%constrain dust within the observed rings and/or ``sweep up'' dust
%grains causing inner gaps and holes, these pronounced morphologies
%would herald the location of extra-solar planets \citep{KB}. Such a
%statement would greatly aid current planet identification
%techniques. Knowledge of the planet's position within the disk would
%also serve as an important verifier of current planet formation
%theories.

Interestingly, \citet{wyatt} argued that the process of grain-grain
collisions can itself act as a potential confining mechanism for the
narrow ring. For most observed debris disks, small dust grains collide
destructively on timescales much shorter than that of P-R drag.
Consequently, dust generated at large distances from the central star
cannot penetrate the collisional barrier to reach the inner region of
the system. This produces a dust ring with a sharply defined
inner-edge. If this dust ring is also optically thick in the radial
direction, so that radiation pressure cannot  disperse the disk (see
above), it is possible for the ring to remain narrow.

%Gurtina Aug 3 : new paragraph 
\citet{KB} introduced an alternative theory for ring formation.
%GB AUG 31 fixed spelling of planetesimal
In a planetesimal disk, runaway growth occurs at successively later
times further away from the star \citep[for an overview
see][]{NagasawaPPV}. \citet{KB} suggested that these newly formed
protoplanets gravitationally stir the left-over planetesimals,
% GB AUG 31 - erodes to erode
inducing violent collisions that erode the planetesimal disk.  As
such, at any given time the disk will appear as a broad ring with a
dark inner hole (or gap). The observed rings herald the locations of
%GB AUG 31 - added and
recent runaway growth and the dark holes where growth has completed, or
shadows where planets have yet to form \citep{KB127}.  
%GB AUG 31 :  reworded  They obtained 
The dust masses and luminosities they obtained are 
comparable to those in the observed
debris disks \citep[see also][]{KB602}. However, if these disks are
optically thin in the radial direction, the ring should not have a
well-defined outer edge since the freshly produced grains will spread
out in a dynamical timescale.
%Furthermore, if planetesimals are
%initially placed in a narrow annulus, \citet{Kenyon} showed that this
%mechanism can reproduce the narrow ring seen in HR 4796A, 
%However, they also ignored the smearing effects of radiation pressure
%and grain-grain collisions.

% considered the role of collisions. He shows that 
%if dust grains destructively collide in a time much shorter than the
%P-R drag timescale (or any other inward migration timescale), as he
%argued is the case for most observed disks, dust generated at a large
%distance from the star cannot penetrate the collisional barrier to
%reach the star. This results in a dust ring with sharply defined
%inner-edge. This appears as a plausible explanation to the ring
%structure seen in many systems.

% wu: new paragraph
\citet{krauss} recently considered a novel effect called photophoresis.
They showed that
%GB AUG 31  removed ``an''
 in optically thin circumstellar disks with gas
densities comparable to that in a minimum mass solar nebula (MMSN),
grains can be pushed outward to a critical radius where the gas
% GB AUG 31 : fixed apostraphe
`photophoretic' force equals the gas drag. The width of the resulting
dust ring is, as of yet, unexplored in this model. Even so, this model
faces the following two difficulties: (1) For the critical radius to
be at 70 AU, the gas density needs to be twice that in the MMSN. This
creates a problem -- there can be no photophoresis effect once the gas
has blocked out the direct starlight. At this density, the optical
depth caused by hydrogen Rayleigh scattering alone is $\tau_{\rm Ray}
= 0.4$ if integrated from 10 AU to 70 AU, and $24$ if integrated from
1 AU to 70 AU
\citep[Rayleigh opacity $\kappa
\sim 10^{-3} \cm^2/\g$ at 4000 \AA,][]{mayer}.  
This model thus requires the gas disk to be truncated inward of $\sim
10$ AU. (2) The assumption of twice the MMSN density for the observed
debris disks has been ruled out by the observed upper limits on the
column densities of multiple gas species (not just hydrogen) in these
disks \citep{Greaves2,Chen}. If we assume a gas density of $2\times
10^{-5}$ MMSN instead (like that in our model), and that grains and
gas are tightly coupled, the critical radius for photophoresis should
be at $\sim 0.2 \AU$
(eq. [13] of \citet{krauss}). This radius is even smaller if grains and
gas are not tightly coupled. For these two reasons, we believe that
photophoresis cannot explain rings in debris disks.
%Two considerations render this
%process irrelevant for producing the observed rings: 1) even without
%dust, a gas disk of such high density would be optically thick due
%to Rayleigh scattering by hydrogen; 2) as the gas disk dissipates to
%form a low density disk, the predicted radius for the dust ring
%becomes too small to be consistent with those observed (cf.
%Table~\ref{tbl-1}).

Along a different line, \citet[][hereafter referred to as {\bf
TA01}]{TA} suggested that rings arise naturally in disks passing
through the transitional phase from gas-dominated (proto-planetary)
disks to dust-dominated (debris) disks. Residual gas in these disks
can be rotating with either super- or sub-Keplerian speeds depending
on the local radial pressure gradient, whereas dust grains have
sub-Keplerian speeds that depend on grain size. The grains therefore
feel either a head-wind or a tail-wind, forcing them to migrate
radially until they are in co-rotation with the gas. TA01 showed that
this effect causes a local concentration of grains; in particular,
when the gas pressure experiences a sharp drop-off, a narrow dust ring
forms locally. In their work this pressure drop-off is provided by an
artificial cut-off in gas density over a radial distance of order the
disk vertical scale height.

In this study, we propose an extension to TA01's work. We lift the
undesirable assumption of an artificially-truncated gas disk and show
that narrow rings form under more general initial conditions. We
assume that the residual gas is smoothly distributed within the disk,
but that the dust is locally enhanced. The enhancement can occur over
%GB AUG 31:  removed pluralization of AU
a distance of tens of AU, e.g., as the result of a major collisional
event in the underlying planetesimal disk (\S
\ref{sec:observemass}). We relax the conventional assumption of
temperature equality between the gas and dust components in the disk
and show that the gas heats up due to the dust enhancement. We
demonstrate that this gas response leads to the confinement of the
dust grains. The mechanism discussed in this paper is an instability:
a slight confinement of the dust heats the gas further which leads to
a stronger confinement.

%\subsection{Comparison with Klahr \& Lin (2005)}
%\label{sec:compare}

\citet{Klahr} (hereafter KL) have proposed a similar mechanism for the 
creation of narrow rings.
%Both works are based upon the same physical idea: in an optically thin
%disk the local gas temperature rises with increasing dust density; the
%resulting pressure gradient closes the positive feedback loop by
%collecting more grains into a narrower region.
Their analytical work assumes a simple relationship between the gas
temperature and dust density, $T_{\rm gas}
\propto \rho_{\rm dust}^\beta$ with $\beta > 0$. Our thermal analysis
validates this assumption for a large range of gas/dust masses and
stellar spectral types. While KL study grains of a single size and
argue that narrow rings will arise from infinitesimally small
perturbations of a smooth dust distribution, we study a continuous
spectrum of grain sizes and a large perturbation (a dust belt).
KL also speculate that the dust ring ``freezes'' after it has formed,
despite the disappearance of the gas as the disk evolves. This
speculation needs to be 
assessed by taking into account radiation
pressure and frequent grain collisions (see above).

In the following, we discuss the observations related to dust and gas
masses in known debris disk systems (\S \ref{sec:observemass}); these
motivate our model for the transitional disk (\S
\ref{sec:fiducial}). In \S
\ref{sec:real}, we present calculations of gas heating/cooling
rates and temperatures in such a disk. Narrow dust rings are shown to
arise in our fiducial model, as well as in a wide range of parameter
% GB AUG 31 added \S
space (\S \ref{sec:instability}). We discuss the best tracers for gas in
these disks (\S \ref{subsec:howto}), as well as major complications to
our model (\S \ref{sec:caveats}). We conclude in \S
\ref{sec:conclusion}.

\section{Dust and Gas masses in Observed Debris Disks}
\label{sec:observemass}

\subsection{Dust Mass}
\label{subsec:dustmass}

Inferred dust masses for various systems are listed in Table
\ref{tbl-1}. Our fiducial value corresponds to that observed in HR
%GB AUG 31 : took mm out of $$
4796A. To produce $\sim 0.1 M_\oplus$ of dust (size $\leq 1$ mm), the
parent bodies (planetesimals or proto-planets) are likely much more
massive. For comparison, the current Kuiper belt has a mass of $\sim
0.1 M_\oplus$,
%wu: I got this out of a website: http://www.plutoportal.net/aboutkuiperbelt.htm
% by Allan Stern
while the primordial Kuiper belt is believed to be many times more
massive (perhaps up to a factor of 50).

To produce the observed infrared radiation, the dust disk must have a
vertical scale height of 
\be
{H\over r} = {1\over 2}{{L_{\rm IR}}\over{L_\ast}}\, 
{\rm Max}\left( {1\over{\tau_r}}, 1 \right), 
\label{eq:Hdust}
\ee
where $L_{\rm IR}/L_\ast$ is the disk infrared to stellar optical
light ratio ($\sim 5\times 10^{-3}$ in the case of HR 4796A
%GB AUG 31: changed the table reference  and H/R to H/r and punctuation
(Table \ref{tbl-1}) and $\tau_r$ is the radial optical depth of the
dust disk. In our fiducial model, we take $H/r = 0.05$.
%wu: this is inconsistent with tau_r = 0.14

The vertical scale height is damped by grain-grain collisions in the
collisional timescale. In our fiducial disk, dust travelling on
inclined orbits will collide with grains at the mid-plane at a
relative velocity of
\begin{equation}
\frac{v_{\perp}}{v_{\rm kep}} \approx {H \over r} = 0.05 .
\end{equation} 
The mean free time for collisions ($T_{\rm collision}$) can be
estimated as
\begin{equation}
% GB AUG 31  : changed coll to collision
T_{\rm collision} = \frac{1}{n_{s}\sigma_s v_{\perp}}
%\hspace{0.5cm} \rm s 
\label{g2},
\end{equation} 
where $\sigma_s = \pi s^2$ is the geometrical cross section of the
grain.  Only collisions between particles of similar sizes are
relevant for the dynamics, thus $n_s$ is approximately the number
density of dust grains in a size bin (s,2s):
\be
n_s = \int_s^{2s} {{dn_{\rm dust}}\over{ds'}} ds',
\label{eq:ns}
\ee
where $dn_{\rm dust}/ds'$ is described by eq. [\ref{eq5}].  
$T_{\rm collision}$, estimated at 70 AU, is
 plotted as a function of
grain size in Fig. \ref{fig:migration} (dotted line). The collision
timescale for the smallest grains is $\sim 20$ orbits in HR 4796A.

The dust vertical velocity can also be damped by gas drag. In the weak
coupling limit, this is described by the following equation,
\begin{equation}
\ddot{z} = \Omega_{\rm kep}^2 \, z - 
{{\rho_{\rm gas} v_{\rm th}}\over{\rho_{\rm grain} s}} \, \dot{z}
\label{g6},
\end{equation}
% GB AUG 31: added  ``is''
where $z$ is the vertical position of the grain and $v_{\rm th}$ is the gas
thermal speed.  We take $\rho_{\rm grain}$, the bulk density of
grains, to be $1.25 \g/\cm^3$.  If this process alone is important,
the vertical scale height of the dust disk decays with a timescale of
$T_{\rm settle}\sim
\rho_{\rm grain} s/\rho_{\rm gas} v_{\rm th}$; this 
%GB AUG 31:  changed   , which  to ; this 
is also plotted in Fig. \ref{fig:migration}.

The frequent collisions and the rapid settling by gas drag should
result in a razor-thin debris disk. The observed finite thickness
therefore indicates stirring \citep{wetherill,kenyon}.

However, giant planets are not necessarily required for the stirring.
For instance, a Kuiper belt with individual bodies $\sim 200$ km in
size, self-stirred to their respective surface escape velocities, and
having overlapping Hill spheres, can maintain a dust scale height of
%GB AUG 31 :  changed H/R to H/r
$H/r \sim 0.05$,
%(the observed lower limit of $H/r \sim 4 \times 10^{-3}$ corresponds to 20 km
% sized objects). 
Dust is continuously wafted up by these bodies. The total number of
such planetesimals required so that their Hill spheres overlap
radially and vertically over a belt-width of 20 AU and scale height of
$H/r = 0.05$ is $\sim 3000 \times 150$, or a total mass of $\sim 2
M_\oplus$ 
% GB AUG 31 : changed reference order
\citep[see also][]{KB121,Goldreich}. The copious amount of dust observed may 
also be produced by collisions between these objects.
%the self-stirring time of this objects may take too long, on the
% other hand, one sees such a hot BO disk

The observed lifetime of such a massive dust disk is related to the
collision time of the largest particles that are in collisional
equilibrium, which is $\sim 1$ Myr for 1 mm particles
(Fig. \ref{fig:migration}).

\subsection{Gas Mass}
\label{subsec:gasmass}

Except for some detections of CO \citep{Roberge,Dent2}, CI and CII
\citep{Roberge, kamp2003,roberge05} and heavy metal absorption and
emission lines \citep{Olofsson,Brandeker}, most gas observations of
debris disks have turned up with only upper limits on the gas column
% Gurtina Aug3 : added the Meyer PPV reference in the review of gas obs. 
density \citep[for a review, see][]{jura-review,MeyerPPV}. These translate to
gas-to-dust ratios of less than a few
%GB AUG 31: changed the order of the references
\citep[][also see \S \ref{subsec:howto}]{Greaves2, 
Lecavelier, Chen, Hollenbach,roberge05-AU}. We set our gas-to-dust
mass ratio to be in the same range.

Gas can be dissipated either via the viscous spreading associated with
MHD turbulence, or by photo-evaporation
\citep[e.g.][]{Alexander}. In the former case, the disk
dissipates gradually with a viscous timescale $\sim 10 $ Myr (for
% GB AUG 31:  changed H/R to H/r
$\alpha = 10^{-2}$ and $H/r = 0.05$ at 70 AU). HR 4796A belongs to the
8 Myr old TW Hydrae association, whose namesake TW Hydra is still
actively accreting. 
% GB AUG 31: reworded below
As such, the assumption of some amount of gas in HR 4796A
does not seem unreasonable.
%surprising.

Photo-evaporation models, on the other hand, predict that once the
inner gas disk has dissipated, which occurs at $\sim 10$ Myr for a 2 
M$_\odot$ star with an ionizing flux of $10^{42} \s^{-1}$ 
\citep[Table 1 of][]{Alexander}, 
%GB AUG 31: changed the order of the reference 
% with an ionizing flux of $10^{42} \s^{-1}$, 
the outer disk disappears in a very short
interval. If this is the case, our assumption of a gas disk will be
difficult to justify. However, we caution that the success of the
photo-evaporation models depends critically on the EUV flux, which is
uncertain even for T Tauri stars. Also, while T Tauri stars (on which
these models are based) are likely magnetically active and bright in
EUV, this may be less true for A-type or F-type stars.  If one adopts
instead a flux of $10^{40}
\s^{-1}$, the photo-evaporation timescale rises by a factor of $\sim 4$
%(extrapolating from their Table 1). 
and viscous evolution may dominate the evolution at the epoch of
interest.

\section{Our fiducial model}
\label{sec:fiducial}

For our fiducial model, we consider an optically thin disk surrounding
an HR 4796A-like star (A0V, $M_\ast=2.5M_\odot, L_\ast =21 L_\odot$,
$T_{\rm eff} = 10,000
\K$). We assume that the disk is in the so-called `transitional phase'
with dust mass comparable to that of the residual gas, although the
underlying planetesimal population that produces this dust can have a
much greater mass.
Our choices for the masses of the gas and the dust components are
justified in \S \ref{sec:observemass}.

The mass of the microscopic dust (grains with size $< 1 \mm$) is taken
to be $M_{{\rm dust}} = 0.1M_\oplus = M_{\rm gas}$. Dust grains are
initially distributed within a broad (FWHM of $20$ AU) density
enhancement centered at $70$ AU. The mid-plane dust density goes as
\begin{equation}
\rho_{{\rm dust}}(r) = 2.7\times 10^{-18}  
\exp \left[- \frac{(r-70)^2}{2\sigma^2}\right] 
\hspace{0.5cm} {\rm g}/\cm^3 \label{eq4},
\end{equation} 
where the standard deviation $\sigma = 20/2.355$ AU. Such a
distribution is motivated by the possibility of a major collisional
event in the underlying planetesimal population.
%wu
%(see \S \ref{sec:discuss} for more details).
%
%Even if such an event
%occurs at a single radius, by the time microscopic dust is generated
%radiation pressure, scattering by larger planetesimals and grain-grain
%collisions would have all conspired to spread the grains over a
%significant radial extent. 
In this article we show that the broad dust distribution described by
% GB AUG 31: changed reference notation to square brackets
eq. \refnew{eq4} will progressively narrow into a sharp ring.
%The motivation 
%for such a distribution comes from considering a major collision event
%in the underlying planetesimal population. 

Initially, dust grains are assumed to follow a size distribution of $dn/ds
\propto s^{-4}$ at every radius, where $s$ is the grain radius. 
This differs from the classical $s^{-3.5}$ distribution
\citep{Dohnanyi}, expected from collisional fragmentation equilibrium,
but is compatible with the result from zodiacal dust analysis
\citep{Fixsen} as well as the spectral fit for HR 4796A \citep{Currie}.  With
such a distribution, grains from each logarithmic size bin contribute
comparably to the total mass, while the smallest grains dominate the
total cross-section.  Taking $s_{\rm min} = 7.7
\mu$m (the radiative blow-out size in the HR 4796A system, see below), and
$s_{\rm max} = 1 $mm, with a grain bulk density of $\rho_{{\rm
grain}}=1.25\g/\cm^3$, we obtain the following number distribution,
\begin{equation}
%wyq: the following equation changed
 \frac{dn_{{\rm dust}}}{ds}(r,s) = 1.0\times 10^{-19}
 \exp\left[-\frac{(r-70)^2}{2\sigma^2}\right] s^{-4} \label{eq5}.
\end{equation}
The radial optical depth for this dust disk is $\sim 0.14$. 
Our calculations take into account the associated attenuation of 
starlight.

The dust temperature can be approximated by the temperature of the
% GB AUG 31: changed eqn reference notation
smallest grain at a given radius \citep[also see eq. 28 of][]{KampB},
\begin{equation}
T_{{\rm dust}} = 282.5 {\rm K} \, 
\left(\frac{L_\ast}{L_\odot} \right)^\frac{1}{5}
\left( \frac{r}{\rm {AU}} \right)^{-\frac{2}{5}}\left(\frac{s_{{\rm
min}}}{\mu \rm m}\right)^{-\frac{1}{5}} \hspace{0.5cm}.
\label{eq2}
\end{equation}
These grains are slightly hotter than the corresponding blackbody
temperature because they are perfect absorbers of stellar photons but
inefficient emitters of their own thermal radiation.  

The ratio between the radiation pressure and the gravity acting on the
grain reads as,
\begin{equation}
%wyq: the following equation changed
\beta = \frac{3L_\ast Q_{\rm PR}}{16\pi GM_\ast c s \rho_{\rm grain}}
= 0.46 \left( {{\micron}\over{s}}\right)
\frac{L_\ast}{L_\odot}\frac{M_\odot}{M_\ast} \label{eq6},
\end{equation}
where $c$ is the speed of light and $Q_{{\rm PR}}\sim 1$ is the
radiation pressure efficiency averaged over the stellar spectrum.
%($Q_{\rm {em}} = 4\pi s/\lambda$). 
The size of the smallest grain is determined by setting
$\beta=0.5$.\footnote{Here, we define the blow-out size ($\s_{\rm min}$)
 to be the size
of the largest grain that escapes to infinity when released on a
circular orbit.} For an A0V star such as HR 4796A, this yields $s_{\rm
min}\approx 7.7\micron$. We assume that grains move only on circular
orbits, which
%GB AUG 31: removed  .This 
limits our study to axi-symmetric features; see \S
\ref{subsec:eccentric} for a discussion on eccentricity damping of the
grains. Under this assumption, the dust orbital velocity ($v_\theta$)
is reduced from the local Keplerian value ($v_{{\rm kep}} =
\sqrt{GM_\ast/r}$) to
\begin{equation}
v_\theta = v_{{\rm kep}}(1-\beta)^{1/2}.
\label{eq9}
\end{equation} 

\begin{figure}
\begin{center}
\includegraphics[scale=0.43]{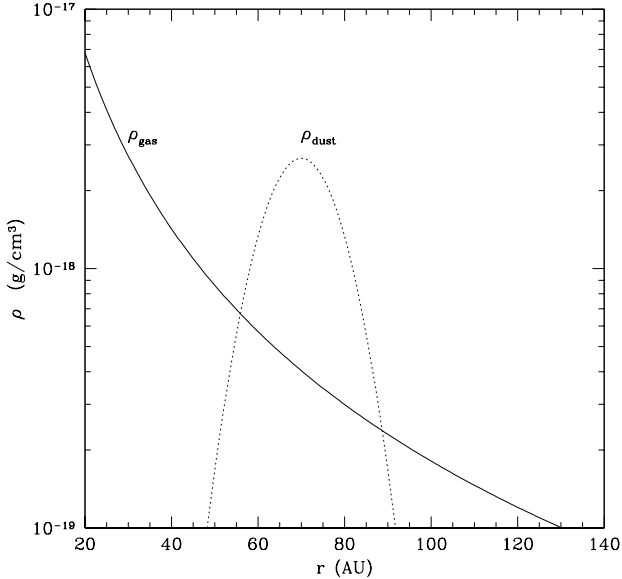}
%\plotone{HR01Density7020.ps}
\caption{
The mid-plane gas and dust densities in our transitional phase
disk. While the gas density (solid line) falls off as a smooth
power-law, the dust density (dotted line) is a Gaussian centered at 70
AU, with a FWHM of 20 AU.  The gas and dust masses are $M_{\rm
gas} = M_{\rm dust} = 0.1M_\oplus$ and both components have the same
vertical scale-height of $H = 0.05 r$. We stress that although we use
these densities for our fiducial parameters, the proposed mechanism
works over a large range of gas and dust masses (\S
\ref{sec:instability}).}
\label{figParam}
\end{center}
\end{figure}

Our gas disk has a total mass of $M_{{\rm gas}} = 0.1M_\oplus$ and
satisfies a mid-plane radial density profile
%using the radial dependence corresponding to TA01's model 1 and
\begin{equation}
%\rho_{{\rm gas}}(r) = 5.7\times 10^{-15}\left ( \frac{{\rm r}}{{\rm AU}}\right)^{-2.25} 
%\hspace{0.5cm} {\rm g}/\cm^3 \label{eq3},
\rho_{{\rm gas}}(r) = 4.0\times 10^{-19}\left ( \frac{{\rm r}}{{70 \rm AU}}\right)^{-2.25} 
\hspace{0.5cm} {\rm g}/\cm^3 \label{eq3},
\end{equation}
where $r$ is the radial distance from the star (measured in AU from
now on). As will become clear, the total mass and density profile
for the gas disk matter little for our theory. What is relevant is the
gas density at 70 AU and the fact that it is smoothly distributed over
a broad region.
We adopt a scale height of $H/r = 0.05$ for the gas
disk, and relate surface density to mid-plane density simply as
$\Sigma = 2 \rho H$. From now on we ignore the vertical structure of
the disk and consider only physical quantities measured at the disk
mid-plane. We also impose an inner and outer cut-off to the disk at 10
and 150 AU, respectively. These cut-offs are far from the region of
interest and do not affect the dynamics. The gas density is plotted as
a function of radius in Figure~\ref{figParam}.

% Parameters

The gas temperature is determined by various heating and cooling
processes (\S \ref{sec:real}).
%Observed stellar properties are summarized in Table~\ref{tbl-1}; $\chi
%= 538$ describes the stellar FUV (11.2 - 12.42 eV) photon flux
% gb: range has changed
The stellar FUV flux (11.2-12.42 eV) is relevant for many gas heating
processes.  Using high resolution stellar spectra provided by Kamp
\& Hauschildt, \citet{Fernandez} obtain a value of $\chi = 538$ for HR
4796A, where $\chi$ is the stellar FUV flux measured at 70 AU,
normalized by the Habing ISM field ($F_{\rm H} = 1.2\times
10^7/\cm^2/\s$ ). This is somewhat larger than the $\chi = 460$
(11.2-13.6 eV) value computed by
\citet{Chen} from FUSE spectra. We adopt the latter value here.

The most important gas heating rate in our disk model is photoelectric
heating by dust grains. The relevant photon energies for this process
range
%For photoelectric heating by dust grains, which is the most important heating
%process in our disk model, the relevant photon energy ranges 
from $W + e\Phi$ to $13.6 \rm eV$, where $W $ is the grain work function 
\citep[$W \sim 8 \rm eV$ for silicate grains, and $\sim 4.4 \rm eV$
for carbonaceous grains,][]{weingartner} and $\Phi$ is the
electrostatic potential of the grain. The value of $\Phi$ depends
(weakly) on stellar spectrum, local electron density and gas
temperature. It is important not to limit the photon energy range to
the FUV range (more below).
%take this flux to be simply the FUV flux (more below).
% and is
%beyond the scope of this work. We take it to be $0$. We parametrize
%the ratio between mid-UV ($W$-11.2 eV) and FUV photon flux as
%$\omega$; $\omega \gg 1$ for both grain types.  For simplicitiy's
%sake, we take $\omega = 1$ in the thermal analysis of \S
%\ref{sec:real}. This substantially underestimates the photoelectric
%heating rate and is the 'worst case scenario' for our
%mechanism. Impact of adopting the realistic $\omega$ value is
%discussed in \S \ref{sec:discuss}.

Gas orbiting around a star is subject to the forces of gravity and gas
pressure. We again assume that gas moves only on circular orbits.  The
gas circular velocity ($v_{{\rm gas}}$) is
\begin{equation} 
v_{{\rm gas}} = v_{{\rm kep}}(1-\eta)^\frac{1}{2} 
%\hspace{0.5cm} \rm {cm/s} 
\label{eq8},
\end{equation}
where 
\begin{equation}
\eta = -\frac{r}{v_{\rm kep}^2 \rho_{\rm gas}}\frac{dP}{dr} \label{eq7}.
\end{equation} 
Here, $P = \rho_{\rm gas} k T_{\rm gas}/\mu_g m_H$ is the gas
pressure, and we take $\mu_g = 2.34$ (for a mixture of molecular
hydrogen and helium).

A grain that is orbiting faster than the local gas ($v_\theta > v_{\rm
gas}$) experiences a head-wind, loses angular momentum and migrates
inward, and {\it vice versa}.  Grains tend to migrate to stable orbits
where they are in co-rotation with the gas, $\eta(r) =
\beta(s)$. This leads to a size segregation, $s = s(r)$.
This is the dynamical basis for our theory.

Let the original surface brightness of a face-on disk be approximated
by the following expression,
\begin{equation}
%wyq: the following equation changed, H --> 2 H
I(r) = \frac{L_\ast}{4\,\pi\, r^2}\, 2 H \int^{1 {\rm mm}}_{s_{{\rm
min}}}
\pi s^2 \frac{dn_{{\rm dust}}}{ds}ds \hspace{0.5cm} 
%{\rm erg/cm}^2/{\rm s}
\label{eq5a},
\end{equation}
where $L_\ast$ is the stellar luminosity and $H$ the dust
scale-height. We set $H/r = 0.05$ as for the gas disk. We have also
taken the grain albedo to be unity and assumed that the disk is
face-on. After radial migration, grains of size between $s$ and $s+ds$
(where $s > 7.7 \micron$) are concentrated into an annulus between $r$
and $r+dr$. Ignoring collisional evolution during this process, the
dust mass within each $ds$ bin is conserved. This yields a new surface
brightness profile:
\begin{equation}
I(r) = {{L_\ast}\over{4\, \pi\, r^2}}\, \pi s^2 \, {{ds}\over{dr}}\,
{{N_s}\over{2 \pi r}},
\label{eq5b}
\end{equation}
where $N_s ds$ is the total number of grains between $s$ and $s+ds$
integrated over the whole disk,
\be
%wyq: the following equation changed
N_s = \int {{dn_{\rm dust}}\over{ds}} 2\pi r' 2H dr'.
\label{eq:Ns}
\ee

\section{Gas Temperature}
\label{sec:real}

Here, we compute gas heating and cooling rates to determine the gas
temperature profile. Much of this analysis is based on the work of
\citet[][hereafter referred as {\bf KvZ01}]{KZ} although we do find some
disagreements with their analysis. Except where noted, we adopt their
expressions to calculate various rates.

%We restrict ourselves to the disk mid-plane and a radial extent of
%10-150 AU from the central star.
%We calculate the resulting corotation radii for all grains, as well as
%the resulting dust surface brightness profile. 
As grains migrate to their stable radii, the gas temperature profile
evolves, modifying the grains' destination. We do not attempt to trace
this behavior by solving the time-dependent equations
self-consistently. Instead, we approach the problem by studying
systems with dust distributions of differing initial FWHM.
%with different initial dust FWHM

%\subsection{Heating and Cooling Rates} 
%\label{subsec:heating}

Results at the disk mid-plane are plotted as functions of radius in
Fig. \ref{fig:heatcool2}. The equilibrium gas temperature is obtained
by demanding that the local heating and cooling rates balance. The
resulting temperature profile dictates the grain migration (\S
\ref{sec:instability}). Unlike KvZ01, we ignore heating from grain
drifting (see below).
%and address it in Appendix \ref{sec:appendix1}.
% added a section about drift heating after the discussion about the 
% photoelectric effect.

%wu: I re-fashioned some equations to make them look nicer...
\subsection{Heating}
\label{subsubsec:heating}

KvZ01 conclude that the relevant gas heating processes in an optically
thin circumstellar disk around A-stars include collisional
de-excitation of H$_2$, photo-dissociation of H$_2$, H$_2$ formation
on grains, gas-grain collisions (when the grain is hotter than the
gas), photoelectric emission from dust grains and cosmic ray heating. 

% GB AUG 31: changed equation to eq.
While KvZ01 used a fitting formula (their eq. [11]) to estimate
the photoelectric heating rate, we opt for a full calculation that
follows the procedure in \citet{draine78} and \citet{weingartner}. Our
results differ markedly from those obtained by KvZ01 (see below).

The photoelectric charging current per dust grain (of radius $s$) is
\begin{eqnarray}
J_{\rm pe} & = & \pi s^2 e \int_{(e\phi+W)/h}^{\nu_{\rm
max}}d\nu\nonumber\, Q_{\rm abs} Y(h\nu) {F_{\nu}\over {h\nu}} \\ & &
\times
\left[\int_{e\phi}^{h\nu-W} f(E,h\nu) dE\right],
\label{eq:Jpe}
\end{eqnarray}
%GB:  probably need to fix the spacing of this equation
where $W$ is the work function 
%GB AUG 31: rearranged the below
%($4.4 \rm eV$ for carbonaceous grains and $8\rm eV$ for silicon grains)
and $\phi$ is the charging potential of the grain. We assume that the
dust disk is composed of $100\%$ carbonaceous grains ($W = 4.4 
\eV$).  Including silicon grains ($W= 8 \rm eV$) does not significantly
change the photoelectric heating rate (see \S
\ref{sec:instability}).  We adopt an overall absorption coefficient of
$Q_{\rm abs} = 1$, a photoelectric yield $Y(h\nu) \approx
(1-W/h\nu)/2$ and describe the electron energy
distribution as
%$f(E,h\nu) = 6/(h\nu-W)\, E/(h\nu-W)\,(1-E/(h\nu-W))$
%GB AUG 31: I rewrote this 
$f(E,h\nu) = 6E/(h\nu-W)^2\, (1-E/(h\nu-W))$ \citep{weingartner}.
%$f$ is called kinetic energy spectrum for the electrons
% GB: modified the below sentence
%We adopt the form for the electron energy distribution as $f(E,h\nu) =
%6/(h\nu-W)\, (E/(h\nu-W)\, (1-E/(h\nu-W))$,
%$f$ is called kinetic energy spectrum for the electrons
%an overall absorption efficient $Q_{\rm abs} = 1$ and a photoelectric
%yield $Y(h\nu) \approx 1/2 (1-W/h\nu)$ \citep{weingartner}. 
$F_\nu$ is the stellar energy flux measured at the grain's location, where
the central star is assumed to be a blackbody of the appropriate
temperature
%$F_\nu$ is the stellar energy flux measured at the location -- taken to be a blackbody of the appropriate temperature
($10,000\K$ for the fiducial
%GB AUG 31: replaced `` standard''
 model).
%GB: below comment about attenuation moved from paragraph discussing electron 
% number density. 
$F_\nu$ is also attenuated by the radial dust optical depth.
% although
%this value is small in our fiducial model ($\sim 0.14$ at the outer
%edge).  
$\nu_{\rm max}$ is the upper frequency cut-off of the stellar spectrum
and is set at an energy of $13.6 \rm eV$.

% GB AUG 31: changed eqn reference notation
The charging potential is obtained by equating eq. [\ref{eq:Jpe}] to the
thermal electron collection current,
\be
J_e = 4 \pi s^2 e s_e n_e \sqrt{{k_B T_{\rm gas}}\over{2 \pi m_e}} 
\left(1 + {{e\phi}\over{k_B T_{\rm gas}}}\right).
\label{eq:Je}
\ee
% GB AUG 31: made below a new sentence
This is valid when the grains are positively charged ($\phi > 0$), 
%as 
which is the case for the gas density of interest.
We take an electron sticking coefficient of $s_e = 1$.  The factor of
$4\pi s^2$ arises here because the grain receives electrons
isotropically, while it intercepts photons only on the side facing the
star (hence the factor $\pi s^2$ in eq. [\ref{eq:Jpe}]). The relevant
photons for the photoelectric effect lie in the UV/FUV range, where
stellar flux decreases ($\sim$) exponentially with wavelength. The net
result is that the charging potential depends weakly (almost
logarithmically) on the electron density ($n_e$), gas temperature
($T_{\rm gas}$) and radial distance from the star
\citep[see][]{Fernandez}. Moreover, each grain is charged to a 
potential that is independent of its size. An easy result to remember
is that $e\phi \sim $ a few $k_B T_\star \gg k_B T_{\rm gas}$ over the
disk.

Short of calculating the full ionization balance in the transitional
disk, we assume that hydrogen and carbon alone contribute to the
electron density ($n_e$). Contributions from molecules and grains are
irrelevant in our low surface density disk (at $70$ AU, surface
density $\sim 10^{-5}\rm g/\cm^2$).  The ionization fraction of
hydrogen is determined by balancing the primary cosmic ray ionization
rate, $\sim 10^{-17}\s^{-1}$
\citep{SpitzerTomasko},
%GB: reference Spitzer and Tomasko 1968 : 6*10^-18?
with the rate of recombination \citep[with recombination coefficient
%wu: added unit to alpha
$8\times 10^{-11}/\sqrt{T_{\rm gas}}\, \cm^3/\s$,][]{Osterbrock}.
% GB: Tielens & Hollenbach quote 1.9*10^-10/T^0.7
% same as Bates & Dalgarno 1962 (also quoted by Clavel et al 1978)
Carbon, with a first ionization potential of $11.2 \rm eV$, is assumed
to be $50\%$ ionized.\footnote{This is found to be true for the
$\beta$-Pic disk \citep{Fernandez}. This fraction rises somewhat
around hotter stars (like HR 4796A), and when the gas density (and
hence the recombination rate) is decreased. We have not attempted to
model this in detail.} Thus,
\begin{eqnarray}
% GB AUG 31: changed italization of H and C and eqn alignment
&n_e&  =  n_e [\rm H] + n_e [\rm C] \nonumber \\
& & = \sqrt{{10^{-17}\over{8\times
10^{-11}}}  T_{\rm gas}^{1/2}\, n[{\rm H_{\rm tot}}]} + 0.5 n[{\rm C}],
\label{eq:edensity}
\end{eqnarray}
where $n[{\rm H_{\rm tot}}]$ and $n[{\rm C}]$ are the number densities of 
hydrogen and carbon nuclei; in our model, hydrogen and carbon
contribute comparably to the electron density.
% optical depth  - GB: I moved the comment about attenuation to the paragraph
% that talks about the stellar flux.

The associated photoelectric heating rate is
\begin{eqnarray}
%GB AUG 31:  moved d nu into the integral 
\Gamma_{\rm pe} & = & \pi s^2 \int_{(e\phi+W)/h}^{\nu_{\rm max}} \,  d\nu\,
Q_{\rm abs} Y(h\nu)  {F_{\nu}\over {h\nu}}
\nonumber \\
& & \times \left[
\int_{e\phi}^{h\nu-W} (E- e\phi) f(E,h\nu) dE\right],
\label{eq:gammape}
\end{eqnarray}
where each ejected electron leaves the grain with a kinetic energy of
$E-e\phi$. 

A simple way to estimate $\Gamma_{\rm pe}$, which also
clarifies the parameter dependences, is to set $\Gamma_{\rm pe} 
\sim J_e$(1eV/$e$). 
%GB AUG 31: changed ev to eV of
In other words, each photo-electron carries $\sim 1$eV of energy out of
the grain towards
heating the gas.  All other factors being equal, two stars
that have photon fluxes differing by orders of magnitude can have
comparable photoelectric heating rates (\S
\ref{sec:instability}).
%wyq: for HR4796A, each electron goes away with ~1ev, for beta-Pic, ~ 0.5ev
%So even though two stars can have photon fluxes that differ by orders
%of magnitude, the charging potential as well as the energy contributed
%by each electron only vary mildly between them. This leads to a
%relative insensitivity of the photoelectric heating rate on the
%stellar spectral type (\S \ref{subsec:criteria}).

When grains span multiple sizes,
\begin{eqnarray}
\Gamma_{\rm pe} & = & \int_{s_{\rm min}}^{s_{\rm max}} ds {{dn}\over{ds}} \pi s^2 
\int_{(e\phi+W)/h}^{\nu_{\rm max}} d\nu\, Q_{\rm abs} Y(h\nu) 
{F_{\nu}\over {h\nu}}
\nonumber \\
& & \times \left[\int_{e\phi}^{h\nu-W} (E- e\phi) f(E,h\nu) dE\right] .
\label{eq:gammape2}
\end{eqnarray} 

Comparing the above heating rate against that of KvZ01, we find that,
for the same parameters, our rates are typically a factor of $20-1000$
times greater.
% at 70AU, kvz/us = 0.044, at 20AU, it is 0.006
If we adopt instead a cooler stellar spectrum, e.g. that of $\beta$
Pic, the difference increases to a factor of $10^3- 10^5$. 
% at 70AU, kvz/us = ; at 20AU, it is 1.7e-5
%wu: said more explicitly below
This difference arises because KvZ01 used a fitting formula that only
%GB AUG 31 added right bracket and reworded
includes the FUV flux ($h\nu \geq 11\rm eV$). This is suitable for the
hotter interstellar radiation field), but in
% In, 
circumstellar environments, the star emits copious amounts of soft UV
photons (between $4.4\rm eV$ and $10\rm eV$) which are much more
relevant for grain charging and photoelectric heating. Similar to our
study, \citet[][appendix C4]{Gorti04} have also considered stellar
flux between $6\eV$ and $13.6\eV$.

% GB ADDED: Comments on drift-velocity heating:
KvZ01 also report that 
%as grains migrate relative to the gas disk, the
the heating rate due to grain drift relative to the gas
% drift velocity heating (i.e., the rate at which grains
%transfer the momentum gained from the radiation field to the gas via
%collisions) 
%much
greatly exceeds the photoelectric heating rate. We find it to be
unimportant in our calculations.
%We disagree with this
%statement, especially in light of the much greater photoelectric
%heating rate we obtain here. 
The enhanced photoelectric heating aside, the drift velocity is also
lower in our calculations because the dust and gas are weakly coupled.
%, aside from the enhanced photoelectric heating. 
A similar conclusion was also drawn by \citet{Gorti04}.

Another important heating rate comes from gas-grain
collisions. From eq. [19] of KvZ01,
\begin{eqnarray}
\Gamma_{\rm gg} &=& 4.0 \times 10^{-12} \,  n[\rm H_{{\rm tot}}] \, \alpha_T  
\, \sqrt{T_{\rm gas}} \nonumber \\ & &\times (T_{\rm dust} - T_{\rm gas})\, 
\int^{s_{\rm max}}_{s_{{\rm min}}} \pi s^2
\frac{dn_{\rm dust}}{ds}ds   
%\hspace{0.5cm} \rm {erg}/{\rm cm}^3/\rm s 
\label{eq26},
\end{eqnarray} 
where $\alpha_T \approx 0.3$ is the thermal accommodation coefficient.
Gas-grain collisions can contribute to either heating or cooling,
depending on the relative values of $T_{\rm gas}$ and $T_{\rm dust}$.

\begin{figure}[t]
\begin{center}
\includegraphics[scale=0.44]{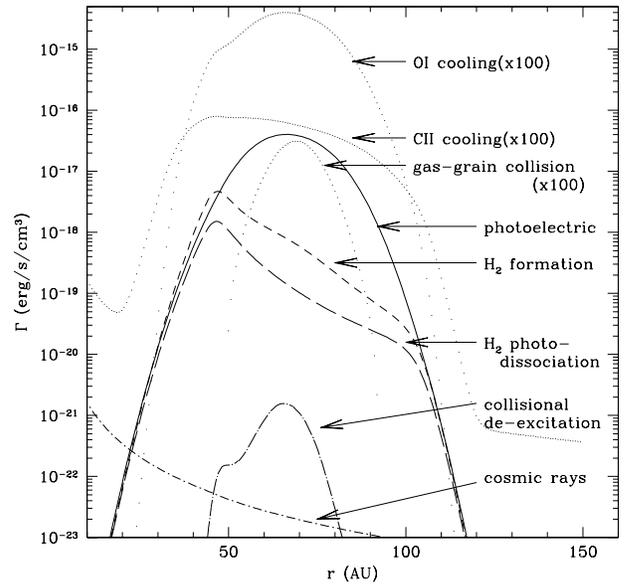}
%\plotone{HR017020HEATCOOLALL.ps}
\caption{
Gas heating and cooling rates are plotted as functions of radius for
the transitional disk around an HR 4796A-like star (A0). The gas is
assumed to be at local thermal equilibrium. All cooling rates have
been increased by a factor of $100$ to distinguish them from the
heating rates. At the peak of the dust enhancement (70 AU),
photoelectric heating and OI cooling are the dominant heating and
cooling processes, respectively. Gas-grain collisions 
%(plotted here using its absolute value) 
act as a cooling mechanism within the vicinity of the dust enhancement
and as a heating rate outside (not plotted). Cosmic-rays dominate the
heating outside the dust belt. H$_2$-related rates are calculated
assuming local chemical equilibrium between atomic and molecular
hydrogen; in the absence of dust, starlight quickly photo-dissociates
all molecular hydrogen.
% --- even a small amount of dust is sufficient
%to drastically raise the H$_2$ formation rate, converting most of the
%atomic hydrogen into its molecular form.
%adding more dust does not increase these rates further. 
The asymmetric shapes of the H$_2$-related rates around 70 AU are
caused by H$_2$ self-shielding of the stellar FUV flux.
%{\bf why does collisional de-excitation has such a form???}  
%In this model, the radially integrated column density for H$_2$ is
%$10^{20} \cm^{-2}$ and $2\times 10^{21} \cm^{-2}$ for atomic hydrogen.
}
\label{fig:heatcool2}
\end{center}
\end{figure}

To calculate H$_2$-related rates (H$_2$ photo-dissociation, 
H$_2$ formation and collisional de-excitation), we need to know the 
ratio of atomic to molecular hydrogen in the disk. This is determined by 
assuming that the H$_2$ photo-dissociation and formation rates are in 
equilibrium. This is a valid approach for a star like HR 4796A, where the 
photo-dissociation timescale ($\sim 10^3$ yrs at 70 AU) is much shorter than 
the system lifetime.  
% not valid for Beta Pic
The relevant stellar photons for H$_2$ photo-dissociation
are in the FUV range ($> 11\rm eV$) and their flux is characterized by 
the value of $\chi$ in Table \ref{tbl-1}.
%value of $\chi$ in Table \ref{tbl-1}.
% GB: we are not using the value in table 1 for HR 4796A
We include self-shielding of the stellar FUV flux by H$_2$ in the
disk, as well as the shielding by dust.
% this is effective once the
%column density of $H_2$ rises above $\sim 3\times 10^{20}$ \citep{Draine} 
The former accounts for the fall-off in the H$_2$ photo-dissociation
rate after the initial rise, seen in Fig. \ref{fig:heatcool2}. 
%$H_2$-related heating are typically less
%important than photoelectric heating or gas-grain collision as they
%are severely attenuated by the $H_2$ self-shielding of far UV
%(1000-1110 \AA) photons \citep{Draine} when even a small amount of
%$H_2$ is present. 
In chemical equilibrium, H$_2$ formation contributes $\sim 4$
times more heating than H$_2$ destruction.
%We take the following simplified approach: if the $H_2$
%photo-dissociation timescale is shorter than the system lifetime, as
%is the case for HR 4796A,\footnote{For an optically thin disk around
%HR4796A, the H$_2$ photodissociation timescale at 70 AU $\sim$10$^3$
%years, much shorter than the age of the system (8$\pm$2 Myr).} one can
%assume that the disk is comprised of atomic hydrogen outside the dust
%enhancement region, while chemical equilibrium exists in the dusty
%region. Equating $H_2$ formation and destruction rates in this region
%(see eqns [17] and [18] of KvZ01) yields the desired ratio between
%atomic and molecular hydrogen. 
% be defined by equating the H$_2$ formation and
%destruction rates (refer to KvZ01 eqns 17 and 18).
%\begin{equation}
%n[\rm H]\, n[\rm H_{\rm tot}]\, R_{\rm form} = n[\rm
%H_2]\, \chi \, \zeta_0 \, f^{\rm H_2}_{\rm shield} \,
%e^{-\tau_d}\label{eq40},
%\end{equation} 
%where $e^{-\tau_d}$ accounts for dust shielding, but in the limit of
%low optical depth this term can be neglected. 
%For young, lower mass stars (for example, $\beta$-Pic), $H_2$
%photo-dissociation timescale is much longer and can exceed the system
%life-time. Chemical equilibrium is appropriate for the dusty region as
%before, but the dust-free region will be occupied mostly by molecular
%hydrogen.

%wyq: we put in pre-shielding before -- removed it in the code now
% as long as setting it to $10^{-2}$ (or smaller, but non-zero), the code 
% doesn't complain 

The conclusion that photoelectric heating is the dominant heating
mechanism in optically thin disks has also been reported by
\citet{jonkheid} and \citet{kamp04}.  
%Its impact on the local gas temperature is more pronounced in our
%transitional disks due to the enhanced dust-to-gas ratio. 
%GB: reworded the below sentence
%photo-electric heating being the dominant mechanism in optically-thin
%disks have also been reported by \citet{jonkheid} and \citet{kamp04}.
%This is more so in our case of enhanced dust-to-gas ratio. 
Furthermore, in contrast to the case of proto-planetary disks, the low
gas density here implies that collisions between gas and grains cannot
cool the gas efficiently. Instead, the gas is heated to temperatures
well above the local dust temperature.

\subsection{Cooling}
\label{subsubsec:cooling}

The low gas density environment considered here is analogous to the
surfaces of protoplanetary disks. Cooling is contributed by atomic
fine structure lines from OI, CII, FeII and SiII, and possibly
molecular transitions from H$_2$, CO, H$_2$O and OH, as well as
% GB AUG 31: changed the order of the references
gas-grain collisions \citep{KZ,Gorti04,kamp04,jonkheid,KD}.  Among
these, we include only the two most important processes: CII and
OI. FeII ($26 \micron$) \& SiII ($35\micron$) can contribute a
comparable amount to the cooling only when the gas temperature exceeds
$\sim 400 \K$. Given the harsh UV environment, CO, H$_2$O and OH are
rapidly photo-dissociated \citep{KampB,Lecavelier} and are negligible
for cooling. In comparison, H$_2$ is abundant since it is continuously
re-formed on grain surfaces.  However, the lack of dipole moment and
the widely spaced energy levels preclude it from being an important
coolant \citep{HM79,Gorti04}.

With a transition wavelength of $158 \micron$, fine structure cooling
by CII is relevant only for gas cooler than $\sim 100 \K$. At hotter
temperatures, the fraction of CII occupying the higher level reaches
unity and the CII cooling rate saturates. OI fine-structure cooling
(dominant line at $63 \micron$) kicks in at these temperatures. This
rate saturates in turn at $\sim 500 \K$.

We assume standard ISM abundances for carbon and oxygen, $n[\rm {C}] =
1.4\times10^{-4}n[\rm H_{\rm tot}]$ and $n[\rm {O}] =
3.2\times10^{-4}n[\rm H_{\rm tot}]$, respectively. Also, following the
results in \citet{Fernandez}, we take carbon (ionization potential of
$11.2\rm {eV}$) to be $50\%$ ionized and oxygen ($13.6\rm {eV}$) to be
$100\%$ neutral. We calculate occupation levels for CII and OI
assuming local thermal equilibrium (LTE); the Einstein coefficients
and statistical weights are listed in Table 1 of KvZ01.  When the gas
density is below a critical value, LTE is inaccurate as the higher
levels may not be sufficiently collisionally populated (for more
details, see KvZ01).  In our model, the gas density at 70 AU ($n[\rm
H_{{\rm tot}}]\approx 10^5 \cm^{-3}$) lies just below the critical
density for LTE. So our faulty LTE assumption overestimates the true
cooling rates and underestimates the resulting gas temperature -- a
more detailed treatment will further strengthen the instability
described here.

We have also verified that these infrared lines are optically thin in
the vertical direction.\footnote{For disks with much higher masses,
these lines can become optically thick. This is accounted for by
reducing the cooling flux by a factor $\tau$.}

When the gas temperature exceeds the dust temperature, gas-grain
collisions act as a cooling mechanism for the gas. In our fiducial
model, this cooling rate is subordinate to line cooling. However, in
denser disks, this cooling process likely dominates.

%where it saturates in turn ($\Lambda_{{\rm OI}}\approx 10^{-16}$
%erg/s/cm$^3$).
%In this regime ($\sim10^{-17}$ erg/s/cm$^3$, $T_{\rm gas}
%\gtrsim 100$ K) . At lower temperatures ($\lesssim$ 40 K)
%$\Lambda_{{\rm CII}}\approx\Lambda_{{\rm OII}}$, 

%Notice also that because $\Gamma_{\rm gg}$ does not dominate, Toy
%Model 2 significantly underestimates $T_{\rm max}$.

\section{The Instability}
\label{sec:instability}
 
\begin{figure*}
\centering{
%\vbox{\psfig{figure=heatcool3.ps,width=1.00\hsize,%
\vbox{\psfig{figure=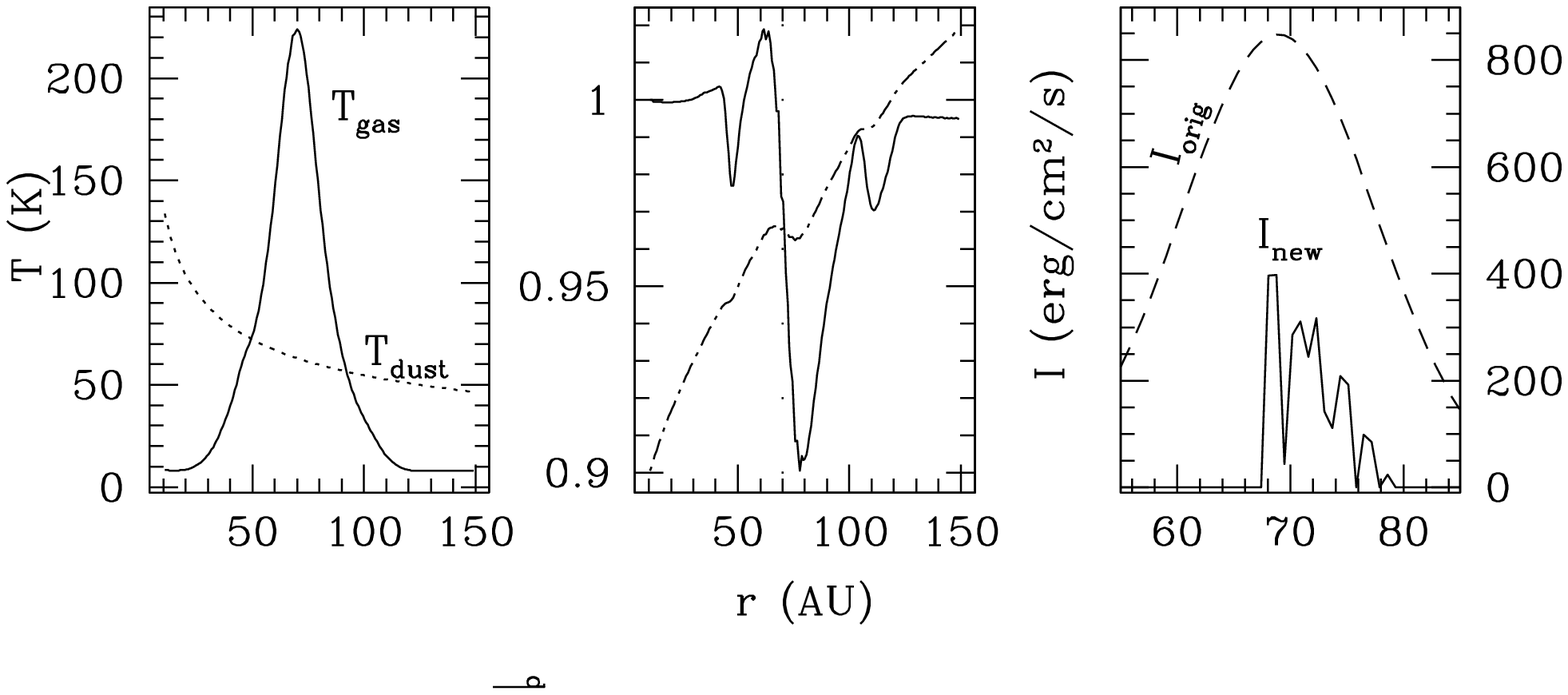,width=1.00\hsize,%
       bbllx=20pt,bblly=160pt,bburx=600pt,bbury=400pt,clip=}}
%par
        }
\caption{Thermal response of the gas disk as in Fig. \ref{fig:heatcool2}.
The left panel shows gas and dust temperatures as functions of radius
(in AU), center panel shows the corresponding gas velocities
%GB AUG 31: changed notation of eqn
(eq. [\ref{eq22}]), normalized by the local Keplerian velocity. The
dotted line marks the centroid (70 AU) of the dust belt; while the
dashed curve indicates the profile of specific angular momentum (in
arbitrary unit). Gas with outwardly decreasing angular momentum
suffers from Rayleigh instability; we expect gas in these regions to
mix rapidly and somewhat reduce the temperature gradient. The right
panel depicts the surface brightness of a face-on disk: the dashed
curve is the original surface brightness profile, while the solid one
is the profile after the grains have migrated to their stable orbits.
The surface brightness profile resulting from dust migration ($I_{\rm
new}$, right panel) is sharply concentrated around 70 AU with a FWHM
of $\sim 10$ AU. The fluctuations seen are numerical artifacts.
\label{fig:heatcool3}}
\end{figure*}

The gas temperature profile obtained by balancing realistic heating
and cooling rates is shown in Fig. \ref{fig:heatcool3}.  Following the
method outlined in \S \ref{sec:fiducial}, we compute the gas and dust
orbital velocities, determine the locations of the stable orbits for
different grain sizes, and calculate the new surface brightness
profile (Fig. \ref{fig:heatcool3}). The latter is more narrowly peaked
around 70 AU, with a FWHM of 10 AU, compared to the 20 AU FWHM of the
initial distribution. More significantly, the new dust belt has both
sharp inner and outer edges.
% gurtinaJune6: reworded the below
%The new dust belt has sharp inner and outer edges.

%The corresponding surface brightness profile is plotted in the centre
%image in Figure~\ref{figBright}. As in the ``toy model'' scenario, the
%resulting dust band has very sharp inner and outer boundaries -
%i.e. the initial broad enhancement has narrowed to a well-defined ring
%of FWHM $\lesssim$ 10 AU centred at $\sim$70 AU. Note that because
%$I(r)$ is a function of the incident stellar flux, the predicted ring
%appears slightly asymmetric (the inner edge is brighter). Furthermore,
%since $\sim9\mu$m-sized grains were not confined, the ring is dimmer
%than in Toy Model 1.

\begin{figure*}
\centering{
%\vbox{\psfig{figure=heatcool3_10AU.ps,width=1.00\hsize,%
\vbox{\psfig{figure=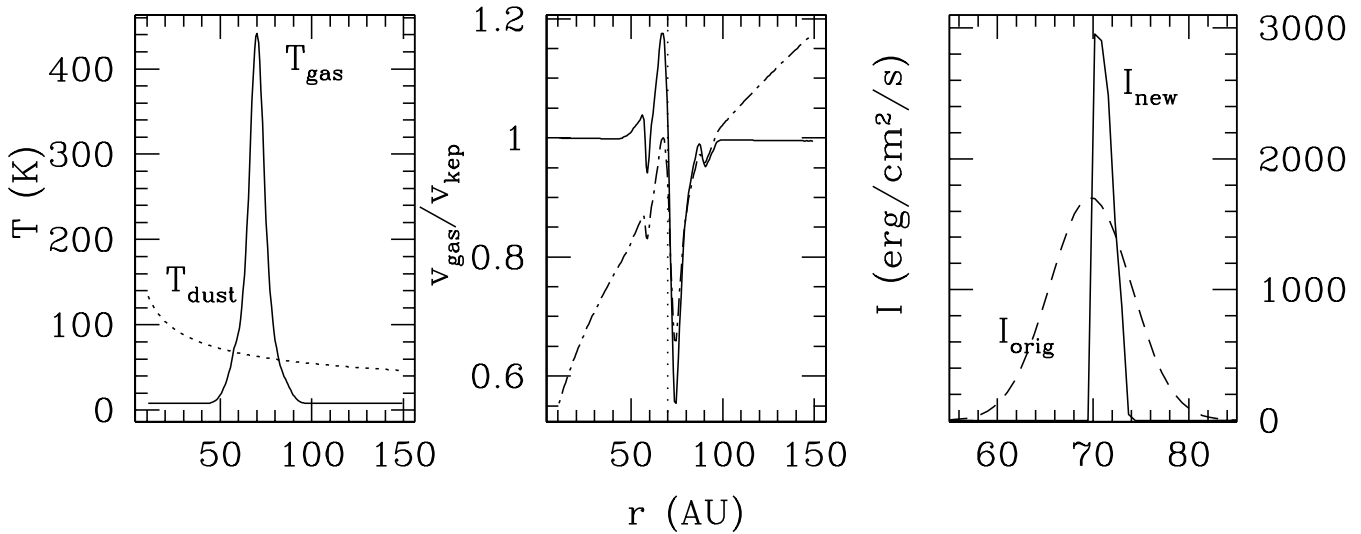,width=1.00\hsize,%
       bbllx=20pt,bblly=160pt,bburx=600pt,bbury=400pt,clip=}}
%par
        }
\caption{Same as Fig. \ref{fig:heatcool3} but the dust is initially
distributed within a narrower region (FWHM = 10 AU). Both the peak
temperature and the maximum $\eta$ reach higher values than those in
Fig. \ref{fig:heatcool3}. The final surface brightness profile is
narrowly confined between 70 and 75 AU -- the ring progressively narrows.
\label{fig:heatcool3_10AU}}
\end{figure*}

\begin{figure*}
\centering{
\includegraphics[scale=0.33]{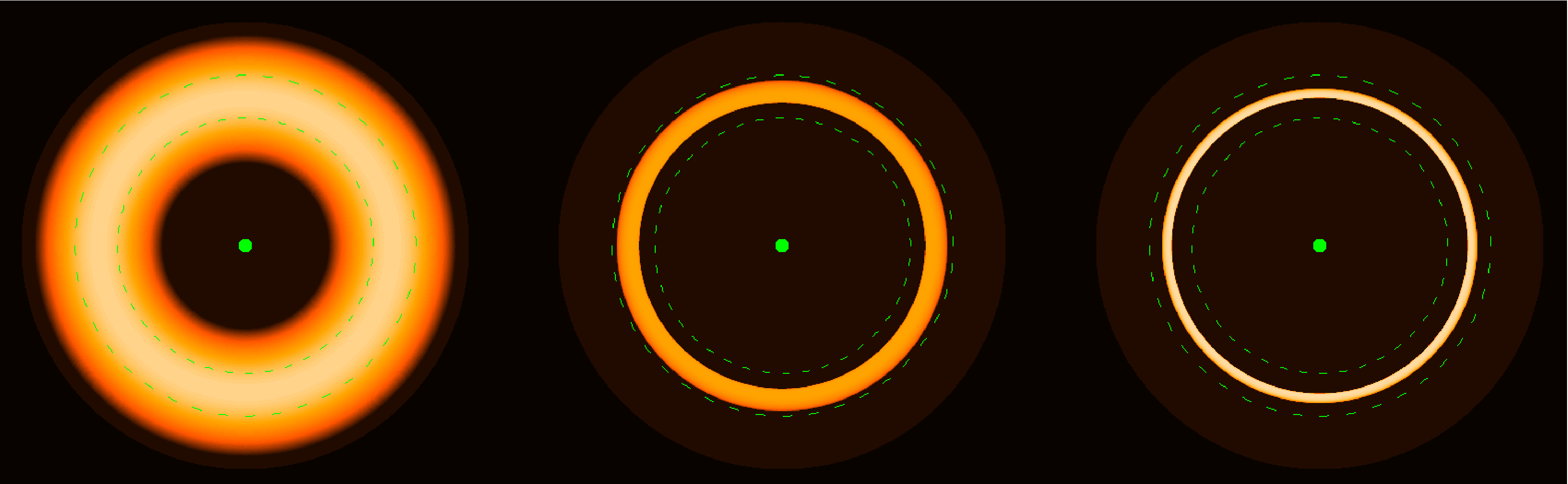}
%\vbox{\psfig{figure=HR.ps,width=1.00\hsize,%
%       bbllx=20pt,bblly=160pt,bburx=600pt,bbury=400pt,clip=}}
%       }
\caption{A graphical demonstration of the evolution of the dust ring,
seen in scattered light. The surface brightness profiles are taken
from Figs. \ref{fig:heatcool3}-\ref{fig:heatcool3_10AU} with the color
white indicating brighter regions. The initial dust enhancement (left)
has a FWHM of 20 AU. The gas within the enhancement heats up and
concentrates grains into a distribution with a smaller FWHM (10 AU,
center image) and sharp inner and outer edges.
% left) is allowed to evolve. After grain migration the
%distribution has narrowed (FWHM = 10 AU, center) and has a sharp inner
%edge which is brighter than the outer edge because of dust
%shielding. The ring is also dimmer than the original distribution due
%to the loss of small grains.
Subsequent evolution of this new distribution produces a bright ring
with a FWHM of 5 AU. We have assumed that small grains are regenerated from
larger grains after being initially lost from the system.}
%progressively narrows and brightens, since
%collisionally regenerated smaller grains are trapped.
%The final
%structure is a bright, sharply defined ring of FWHM of 5 AU.  {\bf Fix
%this caption}}
\label{fig:Bright}}
\end{figure*}

We also perform the same analysis with a narrower initial dust
distribution (FWHM = 10 AU); results are shown in
Fig. \ref{fig:heatcool3_10AU}. The higher initial dust density in the
enhancement raises the photoelectric heating rate and consequently the
gas temperature.  The final surface brightness is strongly peaked
around 70 AU with a FWHM of 5 AU. Although we have not
self-consistently traced the gas temperature evolution as the dust
migrates, the result of this exercise suggests that the segregation
and migration of grains acts as a positive feedback, leading to further
evolution \citep[also see][]{Klahr}. As such, the mechanism we discuss
here is an instability.
% gurtina: reworded the below
%as the grains segregate and concentrate, a positive feedback
%result of this exercise likely indicates that, as grains segregate and
%concentrate, there is a positive feedback that leads to further
%evolution \citep[also see][]{Klahr}. The mechanism we discuss here
%is an instability

The profile of specific angular momentum (dashed line in the middle
panel of Fig. \ref{fig:heatcool3}) decreases outwardly near 70 AU. In
this region, the gas disk will be Rayleigh unstable and mix. As a
result, the pressure profile will likely be somewhat reduced and the
minimum width of the dust ring may be limited to $\sim 10$ AU. 

Two key parameters in this analysis are the gas and dust densities
($n_g$, $n_d$). The photoelectric heating rate scales roughly as $n_g
n_d$,\footnote{The actual rate rises somewhat slower than this because
the electron density increases slower than the gas density
(eq. [\ref{eq:edensity}]). Also, when the gas density reaches MMSN
values, grains become negatively charged. This leads to a Coulumb
barrier that significantly reduces the electron collection current.} 
while the gas line cooling rates scale as $n_g$. So the gas
temperature profile is relatively independent of gas density, but
% GB AUG 31: reworded the below
rises with dust density. In addition, the minimum size of 
%for the confine-able 
confinable grains
($\beta < \eta_{\rm max}$) decreases with higher dust density (also
see above).
%At gas/dust densities $\sim 100$ times greater than our assumed
%values, gas-grain collisions may dominate the cooling with a rate
%$\propto n_g n_d$.

We find that ring formation can occur over a very large range of dust
mass: if we set an artificial limit of $T_{\rm max} = T_{\rm dust}
\sim 60 \K$ for ring formation, the minimum dust mass is $\sim 0.005
M_\oplus$.
% GB AUG 31: changed punctuation in below paragraph
%(by which time heating is increasingly dominated by H$_2$ formation).
The upper end for the dust mass is limited by dust opacity: the dust
disk becomes optically thick for $M_{\rm dust} \geq 0.7 M_\oplus$. The
range of gas mass over which this can occur is equally wide: the
minimum gas mass is set by the constraint that gas is dynamically
important, i.e., $M_{\rm gas} \geq M_{\rm dust}$, and the maximum gas
mass approaches that of the MMSN value. At this value a few effects
set in to diminish the instability including:
%. These include
the optical depth due to
gas opacity, the negative charging potential for the grains and the
increasingly important gas-grain collisional cooling.

Ring formation is effective for an initial dust belt spread out as
wide as 40 AU and is not significantly affected when a background
population of dust (with $1\%$ of the local gas mass) is superimposed.
%The thermal response of the
%gas remains similar except for the presence of a heavily heated inner
%edge, where the stellar FUV flux is absorbed by the H$_2$ that forms
%on the local dust. The consequent attenuation of the FUV flux does not
%affect the rate of photoelectric heating in the belt\footnote{Most of
%the photoelectric heating arises from photons with energies just
%slightly above the work function (4.4 eV for carbon).} and a thin ring
%like that in the standard scenario (case 0) is expected to form.

Other parameters, such as grain composition and spectral type, do not
have a significant impact on our model. 
% GB AUG 31: added Since and changed punctuation
Since the work function for silicate grains is 8eV -- much greater
than the 4.4eV for graphite -- photons capable of striking electrons
out of silicate grains must be more energetic and therefore less
numerous. As a result, the charging potential ($\Phi$) for silicate
grains is typically lower by a factor of $\sim 2$, which leads to a
factor of $\sim 2$ reduction in the photoelectric heating rate (\S
\ref{subsubsec:heating}). Similarly, although stars with later
spectral types output drastically decreasing amounts of UV/FUV flux,
the grain charging potential, the electron density and the kinetic
energy carried away by each electron only vary mildly between
them. Indeed we find that bright, narrow rings may form around stars
with a spectral type as late as K2.

\section{How to find the gas?}
\label{subsec:howto}

The gas mass is a critical unknown in our model. If gas is indeed
present, what is the best way to look for it in debris disks?

In Table \ref{table:3} we calculate some gas observables. We assume
the same fiducial parameters (e.g., gas and dust masses of $0.1
M_\oplus$ as in \S \ref{sec:fiducial}), except for the characteristics
of the central star (distance, luminosity and temperature). We argue
below that in some cases the H$_2$ column densities may not be a good
indicator of gas, while OI and CII fluxes (observable with Herschel)
are excellent tracers.

\begin{deluxetable*}{cccccccc}
\tabletypesize{\scriptsize}
%\rotate
\tablecaption{LTE predictions for various debris disks}
\tablewidth{0pt}
\tablehead{
\colhead{Star/Spec. Type} & 
\colhead{Distance} &
%\multicolumn{2}{c}{Disk Mass ($M_\oplus$)} &
\multicolumn{2}{c}{Column Densities\tablenotemark{a} ($\cm^{-2}$)} &
\multicolumn{4}{c}{Line Fluxes at Earth\tablenotemark{b} ($\erg/\cm^2/\s$)}\\
\colhead{ } &
\colhead{(pc)}&
%\colhead{$M_{{\rm gas}}$} &
%\colhead{$M_{{\rm dust}}$} &
\colhead{N[H$_2$]} & 
\colhead{N[H]} & 
\colhead{OI($44.1\micron$)} &
\colhead{OI($63.2\micron$)} &
\colhead{OI($145.5\micron$)} & 
\colhead{CII($157.7\micron$)}
}
\startdata
HR 4796A(A0) & 67  & $1.0\times 10^{20}$ &$1.9\times 10^{21}$ & 
	$7.4\times 10^{-20}$  & $1.9\times 10^{-14}$ & $3.0\times 10^{-16}$ & 
	$5.3\times 10^{-16}$\\
% HR 4796A is 8 Myrs, 
%Fomalhaut(A3)\tablenotemark{c} & 7.7  & $1.6\times 10^{20}$ & $2.0\times 10^{21}$ & 
%	$1.8\times 10^{-18}$ & $2.9\times 10^{-12}$ & $7.2\times 10^{-14}$  &
%	$5.1\times 10^{-14}$\\
%A3V, 200 Myrs, an interpolation between beta-pic and HR 4796A
$\beta$ Pic(A5) & 19  & $2.3\times 10^{20}$ & $1.7\times 10^{21}$ 
	& $7.7\times 10^{-20}$ & $2.1\times 10^{-13}$ & $3.1\times 10^{-15}$ &
	$6.0\times 10^{-15}$ \\
% beta Pic is 12 Myrs
HD 105(G0) & 40  & $9.3\times 10^{20}$ & $2.7\times
10^{20}$ & $8.9\times 10^{-21}$ & $2.9\times 10^{-14}$ & $3.5\times 10^{-16}$
& $1.2\times 10^{-15}$ \\
HD 107146(G2) & $28.5$  & $6.2\times 10^{20}$& $9.0\times 10^{20}$ & 
$1.7\times 10^{-20}$ & $5.6\times 10^{-14}$& $6.8\times 10^{-16}$ &
$2.3\times 10^{-15}$ \\
%HD 107146 80-200 Myrs
% $\epsilon$ Eri(K2) & $3.2$ & $9.3\times 10^{20}$  
%	& $2.7\times 10^{20}$ &
% $1.3\times 10^{-19}$ &$8.6\times 10^{-13}$  & $5.3\times 10^{-15}$ 
%& $1.0\times 10^{-13}$ \\
% epsilon Eridani: 700Myrs?
\enddata
\tablenotetext{a}{Column densities for molecular and atomic
hydrogen assuming that the disk is edge-on and that molecular and
atomic hydrogen are in chemical equilibrium. The latter assumption may
not be accurate for stars later than A0 where the H$_2$
photo-dissociation timescale is much longer than the system
life-time.}
\tablenotetext{b}{The disk is optically thin in these lines for our assumed parameters. 
Moreover, we assume that all species are in LTE. If the number
densities of colliding particles fall below the critical density for
LTE, the lines fluxes will be greatly reduced.}
%\tablenotetext{c}{In the case of Fomalhaut and
%$\epsilon$ Eri, the stellar ages are so advanced that one does not
%expect $0.1 M_\oplus$ of residual gas; we list them only for
%comparison.}
\label{table:3}
\end{deluxetable*}

\begin{itemize}

\item {\bf H$_2$ Column Densities}

H$_2$ column densities have been traditionally used to place stringent
upper limits on the gas mass. For instance, the non-detection of
absorption in HR 4796A establishes the following $3-\sigma$
upper-limits on the column densities of hydrogen:
%N[CII] $\leq 1.4 \times 10^{14} \cm^{-2}$, N[OI] $\leq 2.8 \times
%10^{15} \cm^{-2}$, N[ZnII] $\leq 2.6 \times 10^{12} \cm^{-2}$, and 
for H$_2$ occupying the two lowest rotational levels, N[H$_{2,J=0}$]
$\leq 10^{15} \cm^{-2}$ and N[H$_{2,J=1}$] $\leq 3.7 \times 10^{15}
\cm^{-2}$ \citep{Chen}. 

The HR 4796A disk plane is inclined by $\sim 17^o$ from our
%GB AUG 31: changed H/R to H/r and put brackets around schneider reference
line-of-sight \citep{Schneider}, i.e., for $H/r = 0.05$, the
line-of-sight passes $\sim 5$ vertical scale heights above the
mid-plane . If the gas at 70 AU is isothermal with $T \sim 200 \K$
(Fig. \ref{fig:heatcool3}) and is in vertical hydrostatic equilibrium,
then the gas scale height is $H_{\rm gas}\sim 23 {\rm AU} \gg H_{\rm
dust} \sim 3.5 \AU$. If the gas at high altitudes has the same
%GB AUG 31: added ``the'' mid-plane
molecular to atomic ratio as that in the mid-plane, the line-of-sight
column density of H$_2$ will be N[H$_2$] = $1.2\times 10^{19}
\cm^{-2}$ in our model. This is much greater than the observed limit.
Does this rule out the presence of a significant amount of gas in the
HR 4796A disk?

The gas thermal timescale at 70 AU (=$nk_BT/\Gamma \sim 20$ yrs for
$T=200$ K, where $\Gamma$ is the heating rate) is short compared to
the dynamical time ($\sim$ 400 yrs at 70 AU),
%GB AUG 31: changed  . As such
so gas rising to high altitudes will fall beyond the influence of the
dust and cool rapidly. The vertical isothermal assumption is therefore
invalid and we do not expect much vertical expansion of the gas
disk. Instead, $H_{\rm gas}
\sim H_{\rm dust} \sim 3.5 \AU$. This reduces the line-of-sight
column density to N[H$_2$] $\sim 7\times 10^{15}
\cm^{-2}$. Furthermore, H$_2$ molecules located well above the disk
mid-plane are destroyed rapidly and the ratio of molecular to atomic
species is
% GB AUG 31: removed ``much''
 reduced compared to that at the mid-plane.
%at a rate of $\gamma_{\rm H_2} \sim 2\times 10^{-8}
%\s^{-1}$, which is much greater than the inverse of the local
%dynamical time (the orbital period or the vertical sound-crossing
%time) $\approx 1/400 {\rm yrs}$.
% kai=460, xi=4.2e-11, assume f_shield = 1.0
%This invalidates our simple 1-D chemical equilibrium calculation and
%indicates that the actual N[H$_2$] can be much lower than the computed
%$3.2\times 10^{19}\cm^{-2}$.
% gurtina: removed the footnote
%\footnote{In fact, a naive estimate will
%yield a column density, for a line-of-sight passing 17$^o$ above the 
%midplane, of N[H$_2$] $\sim 3.2 \times 10^{19} \cm^{-2} \exp\left [{-
%\gamma_{\rm H_2} 400 \rm {yrs}}\right] \, \sim 0$.} 
We therefore conclude that the observed upper limit on N[H$_2$] cannot
yet exclude the presence of a significant amount of gas in the disk.
%--- most of the hydrogen along the line-of-sight may be in atomic form
%and may be very cold. 
The upper limits on other gases \citep{Chen} are less constraining
than that of H$_2$ and are compatible with our model.

Apart from HR 4796A, there is also a stringent upper-limit on N[H$_2$]
in the $\beta$ Pic disk \citet{Lecavelier}: N[H$_2$] $< 10^{18}
\cm^{-2}$. This disk is edge-on, so there is not the same complications in
interpreting the column densities as for HR 4796A. Most likely there
is not a significant amount of gas in the $\beta$ Pic disk
\citep[also see][]{Fernandez}. In corroboration with this conclusion, the 
disk does not exhibit a narrow dust ring.

\item{\bf OI $63.2 \micron$ Flux}

In the transitional disks we study, the gas is heated by
photoelectrons from the dust grains, and cooled mostly by radiation in
%GB AUG 31: added ``the''
the OI $63.2\micron$ line. The estimated line fluxes at the Earth are
listed in Table \ref{table:3}.

All objects should be easily detectable by the PACS instrument in the
upcoming Herschel mission. For instance, PACS should be able to
achieve a $5-\sigma$ detection of the HR 4796A system in a mere 6
minutes.  With a diffraction limit of $\sim 8''$, 
% this is the value at 83micron, the pixle are 9"
Herschel may even spatially resolve some of the close-by systems.
%This may turn out to be a very efficient way of discovering
%circumstellar debris disks.

%The diffraction limit of Herschel ($\sim 4.6''$ at $63 \micron$)
%corresponds to a scale of 300 AU at the distance of HR 4796A, while
%the dust ring (and the centroid of the OI emission) lies at 70 AU
%away from the star. 

%http://www.rssd.esa.int/Herschel/pacs.shtml
% low resolution spectragraph, 57-210 micron, 3x10^{-18} W/m^2

Metallic gas (C, O, Fe, Na...) has been detected in the $\beta$ Pic
disk \citep{Brandeker, roberge05}. This gas may be produced during
collisions of dust grains and contains little or no hydrogen
\citep{Fernandez}. We calculate the OI/CII fluxes expected in such a
hydrogen-poor disk and find them to be comparable to those listed in
Table \ref{table:3}. This is because the electron density is not
significantly reduced when hydrogen is absent from the gas (\S
\ref{subsubsec:heating}). The photoelectric heating rate remains largely 
unchanged from that of a hydrogen-rich disk.  So even though the total
gas mass of the disk is very low,
% disk is very low in total gas mass, 
there is good reasons to expect, and therefore to search for, emission
in fine-structure cooling lines.
%heatcool_metal.f
%We convert our code to calculate thermal structure for a gas that is
%composed of $90\%$ carbon plus $10\%$ oxygen \citep{roberge05}. To
%approximate the observed $\beta$ Pic disk, we assume that the dust has
%a mass of $0.04 M_\oplus$, a FWHM of $60$AU centered at $100$AU, and
%that the gas is cospatial with dust but with a total mass of $0.0023
%M_\oplus$ (to reproduce the observed carbon column densities). 
%We calculate the actual column densities -- but H only increases
%	by a factor ~ 2, so not there...

\item {\bf CII $157.7\micron$ Flux}

%HIFI on Herschel has a sensitivity of 3 h n/K, what does that mean?

Detection of the $157.7\micron$ CII line has been reported for the
$\beta$ Pic disk \citep[using ISO,][]{kamp2003} with a flux of
$1.8\times 10^{-13} \erg/\cm^{2}/\s$ ($4-\sigma$ detection). This is
$\sim 30$ times greater than our computed value of $6.0\times 10^{-15}
\erg/\cm^{2}/\s$ (Table \ref{table:3}) and is surprising.
%The result is 2 times greater if we substitute a bumpy dust
% get 1.75e-14
%distribution by one that is smooth and that follows the gas
%distribution (equal in mass to gas).

% ISO doesn't know about HR 4796A

%$6.8\times 10^{-15}$ if reproduce carbon number density (mass $10^{-4}
%M_\oplus$), carbon cooling already saturated...

Ionized carbon has been observed in absorption with a column density
of N[CII] $\approx 2.5 \times 10^{16}\cm^{-2}$ \citep{roberge05}.
Without knowing the actual gas/dust masses and distributions, We can
obtain an upper limit to the CII flux by assuming that $every$ CII
observed in absorption resides in the excited state and spontaneously
radiates. Assuming a vertical scale height of $0.05$ and no
self-absorption, we obtain a flux of $2.4\times 10^{-14}
\erg/\cm^2/\s$ --- still a factor of $10$ below that reported by
\citet{kamp2003}. More investigation is warranted to resolve this discrepancy.
%This raises the doubt on the reliability of the ISO detection.
%the blackbody approximation to the stellar spectrum underestimates the
%soft UV flux by a factor $10$; 
%that the gas disk is heated by the photoelectric effect to a very high
%temperature and is therefore much thicker than the $0.05$ vertical
%scale height assumed here. 

Looking towards the future, the HIFI instrument on Herschel, with a
high frequency cut-off at $1,910$ GHz ($\sim 157.0 \micron$) and a
superb spectral resolution of $5\times 10^6$, may be capable of
detecting this line in debris disks and provide us with detailed
kinematics of the gas in these disks.

Within the temperature range relevant to this work, the critical
electron number density required for LTE in the OI line is $\sim
10^{5}/\cm^3$, while it is $\sim 10/\cm^3$ in the CII line. 
% this is result from Kyryl 
If the disk density falls below these values, the expected line fluxes
are greatly reduced from our estimates and the gas temperature rises.
Similar values pertain for collisions with hydrogens.

\end{itemize}

\section{Complications}
\label{sec:caveats}

In this section we discuss some complications ignored in our analysis
and their impact on our conclusions.  This include the roles of grain
eccentricities, grain-grain collisions and gas dynamics.
%A more
%comprehensive and self-consistent study is necessary to disclose the
%true amplitude of the instability.

\subsection{Eccentric Orbits of Small Grains}
\label{subsec:eccentric}

Throughout our analysis, we have assumed that grains are on circular
orbits. We examine this assumption here.

\begin{figure}[t]
\begin{center}
\includegraphics[scale=0.44]{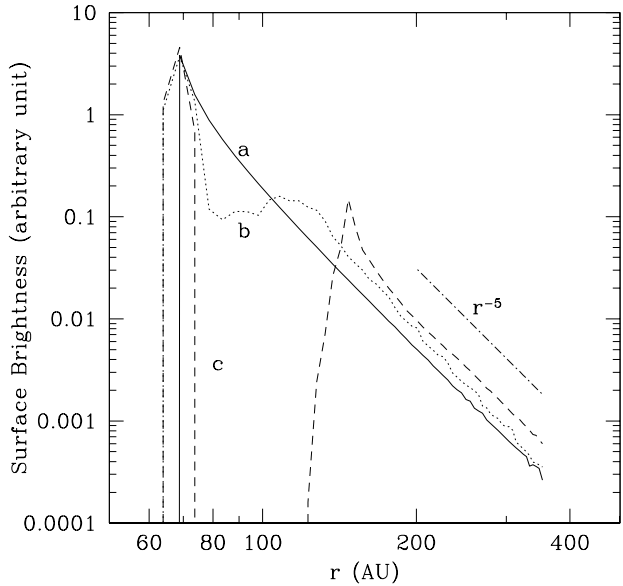}
\caption{Evolution of the surface brightness of a face-on dust disk due to 
gas drag.  Here, dust particles of different sizes ($s^{-4}$ number
distribution) are
% GB AUG 31: replaced all started 
initially produced at 70 AU with the local Keplerian velocity. The
surface brightness is plotted (in arbitrary unit) as a function of
radius (in AU): (a) when no gas is present (or at time $0$); (b) after
undergoing gas drag for $10^4$ yrs; (c) for $10^5$ yrs.  Case (c) will
be observed as a narrow ring with well-defined edges, confirming the
results obtained assuming circular orbits
(Fig. \ref{fig:heatcool3}). The adopted gas parameters are from our
fiducial model (temperature as shown in Fig. \ref{fig:heatcool3}). The
dot-dashed line indicates a $r^{-5}$ power law, confirming the result
in eq. [\ref{eq:sbprofile}] for the gas-free case.}
\label{fig:SB}
\end{center}
\end{figure}

If grains are produced from larger bodies that are on circular
orbits, they would acquire an eccentricity at birth,
\be
e = e_{\rm birth} = {{\beta}\over{1-\beta}} = 
%mistake
{{s_{\rm min}}\over{2 s - s_{\rm smin}}},
%{s \over{2 s_{\rm min}-s}},
\label{eq:graine}
\ee
where numerically, $e \sim (s/s_{\rm min})^{-5/3}$ for $s \sim s_{\rm
min}$. Grains are then launched to highly eccentric orbits with their
%GB AUG 31: reworded and changed punctuation
apastron sorted by their sizes; the dust ring consequently spreads out
in the dynamical timescale. Assuming a grain size distribution of
$dN_s/ds \propto s^{-\alpha}$, where $N_s$ is defined in
eq. [\ref{eq:Ns}], and that grains reside only at their apastron, $r
\approx r_{\rm ap} = 70\times(1+e)/(1-e)$ AU $\sim 140/(1-e)$ AU (for
$e \sim 1$), we obtain the following surface brightness profile
(eq. [\ref{eq5a}]):
\be
I(r) \propto {{s^{2-\alpha}}\over {r^3}}{{ds}\over{de}}
{{de}\over{dr}}
\propto {1\over{r^5}} (1-{{140}\over{r}})^{-{14/5}+3\alpha/5} 
\propto {1\over{r^5}}.
\label{eq:sbprofile}
\ee
% GB AUG 31: added profile and broad
This profile is insensitive to the value of $\alpha$ (see
Fig. \ref{fig:SB}).
%(also see \S \ref{subsec:eccentric}).
Without gas drag this broad surface brightness profile does not evolve with
time and conflicts with the appearance of a narrow ring \citep[also
see][]{strubbe}.

The presence of gas changes this scenario: 
%GB AUG 31: changed the below
%The presence of gas changes this surface brightness profile. For one thing, 
grain eccentricities are typically damped by collisions with gas
particles (which are assumed to be on circular orbits). A grain moving
at a velocity ${\bf
\Delta v}$ relative to the gas feels a drag force
\citep{kwok}
\be
{\bf F}_{\rm drag} = - \pi s^2 \rho_{\rm gas} (v_{\rm th}^2 + \Delta
v^2)^{1/2} {\bf \Delta v},
\label{eq:fdrag}
\ee
where $v_{\rm th} = 4/3 \sqrt{8 k_B T_{\rm gas}/\pi \mu m_H}$ is the
local thermal speed. The relative velocity (as well as the
eccentricity) can be damped in $\sim T_s$ orbits, where $T_s$ is the
dimensionless stopping time (also called the Stokes number):
\begin{equation} T_{\rm s} = t_s \Omega_{\rm kep} =
{{4 \rho_{\rm grain} s v_{\rm kep}}\over {3\rho_g r v_{\rm th}}}
{1\over{\sqrt{1 + {\Delta v^2/v_{\rm th}^2}}}}.
\label{eq21}
\end{equation} 
Here $\Omega_{\rm kep}$ is the local Keplerian frequency and 
% GB AUG 31: reworded the below
%the dimensional stopping time 
$t_s = m_{\rm grain} \Delta v/F_{\rm drag}$ is the dimensional
stopping time, with $m_{\rm grain} = 4\pi/3 \rho_{\rm grain} s^3$. In
our fiducial disk, the gas density is so low that all grains are
weakly coupled to the gas: $T_s$ ranges from 4 for the smallest grains
to $600$ for 1 mm grains. Even so, the dimensional stopping time is
still short compared to the collision time. So large grains can be
assumed to be placed on circular orbits after they are produced.  Very
small grains (near blow-out size, $s\geq s_{\rm min}$), however,
behave differently.

% GB AUG 31: yanqin, for ``very small grains'' you mean near blow-out size? 
The highly eccentric orbits ($e_{\rm birth} \sim 1$) of these very
small grains keep them away from the denser inner gas disk for most of
their orbits.  As a result, the effective stopping time is
longer. Moreover, these grains encounter a tail wind near their
apastron. For very small grains, this positive torque can exceed the
negative torque that they may receive at periapstron.
% GB AUG 31: should it be periapsis here or periastron ? is the closest
%point to the disk midplane also the closest distance to the star? 
They are then migrated outward in addition to being gradually
circularized.  If the gas disk has an outer edge, the eccentricities
of these grains may
%eccentricities for the very small grains may
actually increase (TA01). So it is unreasonable to assume these small
grains follow circular orbits with semi-major axes near 70 AU (the
production site).

For the temperature profile in Fig. \ref{fig:heatcool3}, the
large-small grain boundary is at $\sim 3.1 s_{\rm min}$,
% use eccentricity_damping_singlesize.f to check orbital evolution
while for that in Fig. \ref{fig:heatcool3_10AU}, the boundary is at
$\sim 2.4 s_{\rm min}$. In other words, as the gas temperature
rises due to the increasing dust
%GB AUG 31: reworded : with the dust
concentration, smaller and smaller grains will be retained within the
ring (an instability).

In Fig. \ref{fig:SB}, we show the surface brightness profile of the
dust ring with and without gas drag. Particles of various sizes are
produced at 70 AU with the local Keplerian speed. Their dynamics then
evolve under the combined forces of gravity, radiation pressure and
gas drag (where applicable). In the presence of gas, the dust profile
progressively evolves into a
%GB AUG 31: changed ``more and more'' added progressively 
sharply defined ring.  Note that the local pressure maximum produced
by dust heating is essential for containing grains -- if the gas has
zero pressure gradient or if the gas temperature equals the local dust
temperature, all grains are gradually pushed outward.

% produced with eccentricity_damping.f

Lastly, circularization of bigger grains also reduces the birth
eccentricity of the small grains -- for instance, since the $20
\micron$ grains are circularized by gas before they collide to produce
the $10 \micron$ grains (see Fig. \ref{fig:migration}), the birth
eccentricity for the latter group is reduced from the value in
eq. [\ref{eq:graine}] to
\be
% the following is not an accurate solution, 
%the real one (1+e)/(1-e) = (1-beta/2)/(1-beta)
e_{\rm birth} \sim {{\beta/2}\over{1-\beta/2}}.
\label{eq:ebirthsmall}
\ee
This moves the aforementioned large-small boundary to 
an even smaller value. 

These positive feedback effects may in the end lead to trapping of
even the smallest grains -- an impossible feat when no gas is present.
%GB AUG 31 : added punctuation
A more complete analysis than that performed here is required, but the
presence of non-circular orbits does not seem to change our basic
conclusions about narrow-ring formation.

\subsection{Grain Collisions}
\label{subsec:collision}

%show that small grains in the disk will reach their
%stable orbit location unhindered by collisions. Larger grains,
%however, are expected to encounter many collisions and may regenerate
%the smallest grains that are initially lost to the system.

\begin{figure}[t]
\begin{center}
\includegraphics[scale=0.44]{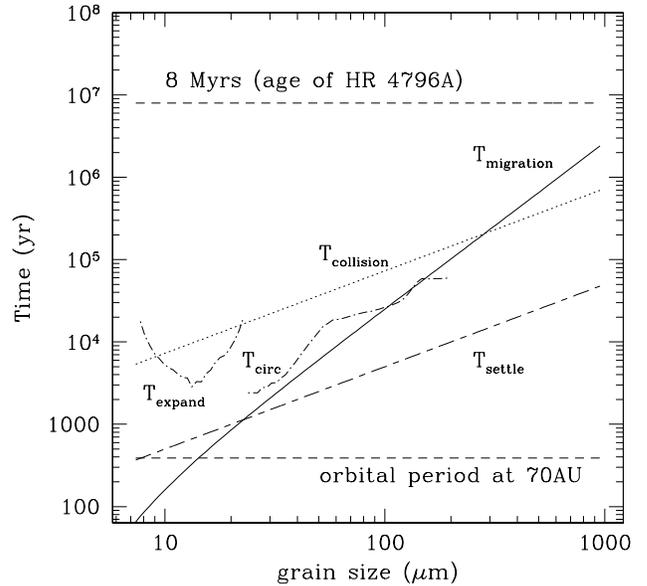}
%\plotone{HR01CollTimeStab7020.ps}
\caption{Various timescales in our fiducial model are plotted here as functions
of grain size. $T_{\rm migration}$ is the radial migration timescale
for grains to move by half the original FWHM of the dust belt.
$T_{\rm collision}$ is the collision timescale between grains of
comparable sizes in the belt. $T_{\rm settle}$ is the vertical
settling time for a grain due to gas drag. The two dot-dashed curves
represent
%GB AUG 31: added punctuation
$T_{\rm circ}$ and $T_{\rm expand}$, which are relevant for eccentric
grains (\S \ref{subsec:eccentric}): the former is the timescale for
gas drag to damp the grain birth eccentricity, while the latter is the
timescale for the periaps of small grains to be pushed outward of,
say, 100 AU, by which distance they
%GB AUG 31: changed their to they
can no longer affect the narrow ring appearance. The dashed lines
denote the estimated age for the HR 4796A system and the orbital
period at 70 AU, respectively. The Poynting-Robertson drag timescale
is comparable to the system lifetime even for $10\micron$ grains, and
is much longer for larger grains. We set $\eta= 0$ for simplicity when
calculating $T_{\rm migration}$.
%and obtain values for the gas density and temperature at 70 AU from
%Fig. \ref{figParam} \& Fig. \ref{fig:heatcool3}. 
%Grains smaller than $\sim400\micron$ reach their stable orbits before
%undergoing a collision.
}
\label{fig:migration}
\end{center}
\end{figure}

We have so far ignored the role of grain regeneration by
collisions. The importance of collisions can be best studied by
comparing different timescales. In Fig. \ref{fig:migration}, we plot:
the orbital period at 70 AU; the timescale for grain collision
($T_{\rm collision}$) in the initial 20 AU FWHM dust belt; the
%GB AUG 31 changed punctuation
timescale for a grain to reach its stable orbit ($T_{\rm migration}$);
the timescale for the grains to vertically settle due to gas drag
($T_{\rm settle}$); the timescale for the grain's birth eccentricity
to be halved by gas drag ($T_{\rm circ}$); and the timescale for the
periapstron of small grains to be expanded outwards of 100 AU
%GB AUG 31 added brackets
($T_{\rm expand}$). The latter two are relevant for eccentric
grains. We adopt the fiducial model (Fig. \ref{fig:heatcool3}) when
calculating these values.

Numerically, we quantify $T_{\rm migration}$ as the time it takes for
a grain originally located at 70 AU to travel a distance of 10 AU
(FWHM/2). The radial migration velocity is calculated as follows. in
the weak coupling limit, the tangential movement of a grain is not
significantly modified by the gas drag, $v_{\theta} = v_{\rm
kep}(1-\beta)^{1/2}$.
%, and the relative velocity between gas and dust is
%\be
%\Delta v \approx v_\theta - v_{\rm gas} = 
% v_{\rm kep}[(1-\beta)^{1/2} - (1-\eta)^{1/2}].
%\label{eq:dv}
%\ee
% The difference between the grain's azimuthal velocity and the
%circular velocity of the gas (eq. [\ref{eq8}]) dominates the relative
%velocity,
% different from the gas circular velocity (eq. [\ref{eq8}]). This difference dominates the relative velocity,
The gas drag associated with the gas-grain relative velocity removes
(injects) angular momentum from (into) the dust grain if the gas moves
slower (faster) than the dust, causing a grain of size $s$ to migrate
to a radius $r$ where $\beta(s) = \eta(r)$.\footnote{The
Poynting-Robertson drag is insignificant in comparison to the gas
drag.} The grains migrate radially with a velocity (cf. TA01
eq. [23])
\begin{equation} 
v_{\rm rad} = \frac{2}{T_{\rm s}}
%\left[1 + \left(\frac{\Delta v_\theta}{v_T}\right)^2\right]^\frac{1}{2}
\left[\left(\frac{1-\eta}{1-\beta}\right)^\frac{1}{2}
-1\right]v_{\rm kep}.
\label{eq22}
\end{equation} 

Circular grains ( $s \geq 300\micron$) typically encounter multiple
collisions -- and possibly grind down to smaller grains -- before they
reach their respective stable orbits. This allows small grains to be
regenerated continuously even if they are initially lost from the
system due to gas drag.
% a requisite for forming bright, narrow rings in some of our models.
Grains smaller than $\sim 300 \micron$ can migrate
%GB AUG 31: changed be migrated 
to their stable orbits before colliding
destructively with grains of a similar size.

%GB AUG 31: changed notation of eqn reference
When grain eccentricities are considered (eq. [\ref{eq:graine}]), all
grains greater than $\sim 20 \micron$ will be trapped in the ring and
are circularized faster than their collision lifetime -- their
collisional progeny will likely obtain a birth eccentricity closer to
that in eq. [\ref{eq:ebirthsmall}] than that in eq. [\ref{eq:graine}],
and are therefore easier to trap in the ring.  Grains of size $\sim 10
- 20 \micron$\footnote{When the gas temperature is raised further due
to the enhanced dust concentration, these sizes are reduced to smaller
values.} are depleted from the ring faster than they can collisionally
break-down ($T_{\rm expand} \ll T_{\rm collision}$) -- or more
relevantly, they are removed faster than they can be regenerated from
larger grains. So these grains will not destroy the narrow ring
appearance. Very small grains ($7-8
\micron$) are removed more slowly. However, their very large apastron also
means that they do not matter much for the narrow ring appearance. 

In summary, grain collisions do not seem to adversely affect our
theory.

\subsection{Gas Dynamics}
\label{sec:gas}

We have ignored all dynamical responses of the gas to the heating and
the migration of the dust. The gas may expand vertically due to the
heating, and expand radially as the dust is concentrated. Moreover,
such an optically thin disk most likely experiences a macroscopic
viscosity due to MHD turbulence. How does gas dynamics impact the ring
instability?

In \S \ref{subsec:howto} we argue that there is little vertical
expansion of the gas within the heated regions of the disk.
%GB AUG 31: changed As to because 
Because the gas is heated or cooled faster than the vertical sound
crossing time, gas
%GB AUG 31 added ``the''
temperature at every altitude is determined by the local heating and
cooling. Gas high above the mid-plane is heated to a temperature lower
than that at the mid-plane,
%GB AUG 31 as changed to since and added thus
since the photoelectric heating yield there is lower.\footnote{This
conclusion can reverse when one considers NLTE cooling.} The overall
vertical scale height is thus determined, not by the mid-plane
temperature, but by that of the dust disk. This justifies our fiducial
choice of setting $H/r$ of the gas and dust components to be equal.
%being equal between the gas and the dust components.

As dust is concentrated into a narrow ring,
% by action-reaction 
the gas is expelled away from the ring by action-reaction, 
which lowers the local gas density and
reduces the pressure gradient. This negative feedback only occurs
because grains are being concentrated. It is a higher order effect
that acts to limit the amplitude, but not change the sign, of the
instability.

Viscosity tends to erase the (angular) velocity gradient and dissipate
the gas disk. However, if the ring is $\sim 10$ AU in width, the
viscous time across the ring is $\sim 0.5$ Myr (taking a
Shakura-Sunyaev $\alpha=10^{-2}$, $H/r = 0.05$). As a result, the
velocity gradient can be actively maintained by dust heating. Gas
dissipation occurs over a few Myr timescale and may eventually
terminate the ring instability.

\section{Conclusions}
\label{sec:conclusion}

We have studied the thermal and dynamical consequences of an optically
thin dust disk embedded in a gas disk of comparable mass. We showed
that photoelectric emission from the dust grains heats up the gas and
this modifies
%GB AUG 31:  Yanqin you mean modifiy the grains' orbital motion right?
%its 
the grains' orbital motion: if the dust is produced in a belt-like
region (like that of our Kuiper belt), the heated gas will collect
dust particles into a narrower region with sharp edges. This resulting
dust distribution may be associated with the dust rings observed in
various debris disks, such as that around HR 4796A, and possibly
%that around 
HD 105.
% GB: removed this footnote as this is pointed out at the end of the previous section
%\footnote{We consider in \S \ref{subsec:constraints} the implications of the observed upper limit in N[H$_2$] from this system.}

%The key physics of this article are presented in
%Figs. \ref{fig:heatcool2}- \ref{fig:heatcool3_10AU}.

This mechanism operates around stars with a spectral type at late as
K2 and is valid for gas and dust masses over $\sim 2$ orders of
magnitude in range. So, if every circumstellar disk goes through a
transitional stage as we have described here, a narrow dust ring is
expected to form. However, it is likely that the ring will disperse as
the gas dissipates.

%4) big grains may be sitting at where they are produced, while small
%grains are trapped by gas

%The morphology of a debris disk may yield a wealth of information
%about the system's evolutionary stage and disk composition. 
% GB AUG 31: changed punctuation and reworded below
We have restricted ourselves to consider only axi-symmetric rings; 
%it is 
an interesting extension of this work will be to explore whether
non-axisymmetric features may also derive from gas-dust interaction.
We have also considered
%GB AUG 31 : added
the impact of eccentricities
%produced 
induced by radiation pressure on small grains that are released from
larger bodies on circular orbits. This process complicates our theory,
but does not diminish the instability.
%Furthermore, perturbations by planets or
%companions can contribute to the observed asymmetric features, such as
%warps \citep[$\beta$ Pic,][]{Mouillet,heap}, clumps 
%\citep[Vega,][]{Wilner} and off-centered rings \citep[Fomalhaut,][]{kalas2}.

%GB AUG 31: added
We conclude that the presence of narrow dust rings in optically thin,
transitional-stage disks, such as that observed around HR 4796A, does
not necessarily imply the presence of ``shepherding'' planets.
Instead, we argue that the detection of narrow dust rings may indicate
the presence of gas. We have shown that this conclusion is compatible
with the stringent upper limit on
%GB AUG 31 added the
the H$_2$ column density in the HR 4796A disk.

Lastly, most of the gas cooling occurs via the OI or CII infrared
lines. We have computed the expected fluxes for various systems and
point out that the upcoming Herschel mission will be instrumental in
observing and characterizing the gas.

%The predicted ring features have sharp inner and outer
%boundaries. However, grain-grain collisions will regenerate $<$
%blow-out sized grains, which are lost to the system within dynamical
%timescales. As such, the outer boundary is expected to smear out as
%the ring evolves, whereas the inner boundary will remain
%well-defined. Furthermore, gas dissipation is expected to induce a
%milder pressure gradient and consequently broaden the ring. Thus
%sharply-defined, bright ring structures may form preferentially
%between the primoridal and debris disk stages of evolution - i.e. in
%transitional-stage disks.

%Consequently, the gas pressure drops sharply without artificially
%introducing a sharp gas-disk edge, as TA01 have done. The grains
%subsequently migrate to stable orbits sharply confined within the FWHM
%of the initial enhancement. Most orbits concentrate at the peak
%density location and the broad distribution narrows to roughly half
%its initial FWHM. A second order analysis reveals that, as the broad
%enhancement narrows, $\Gamma_{\rm pe}$ further increases the local gas
%temperature and induces a steeper pressure drop. As a result, smaller
%and smaller grains are confined within a narrower region and the ring
%is brightens.

\acknowledgements We thank an anonymous referee for 
careful readings and many constructive comments. Thanks are also due
to Alexis Brandeker, Ray Jayawardhana, Rodrigo Fernandez
%providing multiple stellar flux calibrations 
and Kyryl Zagarovky for help and discussions. This research was
supported by the National Science and Engineering Research Council of
Canada, through a discovery grant to YW and an NSERC undergraduate
student research award to GB.
% thank the referee

%\bibliographystyle{apj}
%\bibliography{astro}

\end{document}